\documentclass[11pt]{article}
\pdfoutput=1
\usepackage{jcapmod}

\usepackage{shorthand}
\usepackage{mathtools}
\usepackage{tikz}
\usetikzlibrary{decorations.pathmorphing}
\usepackage{booktabs}
\usepackage[english]{babel}
\usepackage{amsmath,amssymb,amsbsy,amstext, amsthm, simplewick, amsfonts}
\usepackage{hyperref}
\usepackage{graphicx}
\usepackage[small]{caption}
\usepackage{siunitx}
\usepackage{upgreek}
\usepackage{framed}
\usepackage{wrapfig}
\usepackage{multirow}
\usepackage{bbm}
\usepackage[svgnames,dvipsnames,x11names]{xcolor}

\usepackage{selinput}

\usepackage{bm}
\usepackage{float}
\usepackage{geometry}
\usepackage{yfonts}
\usepackage{caption}
\usepackage{subcaption}
\usepackage{sidecap}
\usepackage{longtable}
\usepackage{anyfontsize}
\usepackage{dsfont}
\usepackage{tikz}
\usepackage{relsize}
\usepackage{tcolorbox}

\usepackage{xcolor}
\usepackage{xparse}

\usepackage{slashed}
\usepackage{simpler-wick}

\NewDocumentCommand{\colornucleus}{omme{_^}}{%
  \begingroup\colorlet{currcolor}{.}%
  \IfValueTF{#1}
   {\textcolor[#1]{#2}}
   {\textcolor{#2}}
    {%
     #3
     \IfValueT{#4}{_{\textcolor{currcolor}{#4}}}
     \IfValueT{#5}{^{\textcolor{currcolor}{#5}}}
    }%
  \endgroup
}

\usepackage{array}
\newcolumntype{L}[1]{>{\raggedright\let\newline\\\arraybackslash\hspace{0pt}}m{#1}}
\newcolumntype{C}[1]{>{\centering\let\newline\\\arraybackslash\hspace{0pt}}m{#1}}
\newcolumntype{R}[1]{>{\raggedleft\let\newline\\\arraybackslash\hspace{0pt}}m{#1}}

\usepackage[framemethod=default]{mdframed}
\newmdenv[skipabove=7pt,
skipbelow=7pt,
rightline=false,
leftline=false,
topline=false,
bottomline=false,
backgroundcolor=gray!10,
linecolor=gray,
innerleftmargin=5pt,
innerrightmargin=5pt,
innertopmargin=5pt,
innerbottommargin=5pt,
leftmargin=0cm,
rightmargin=0cm,
linewidth=4pt]{eBox}
\newmdenv[skipabove=7pt,
skipbelow=7pt,
rightline=false,
leftline=false,
topline=false,
bottomline=false,
backgroundcolor=gray!10,
linecolor=gray,
innerleftmargin=5pt,
innerrightmargin=5pt,
innertopmargin=-5pt,
innerbottommargin=5pt,
leftmargin=0cm,
rightmargin=0cm,
linewidth=4pt]{eBox2}
\usepackage[framemethod=default]{mdframed}
\newmdenv[skipabove=7pt,
skipbelow=7pt,
rightline=true,
leftline=true,
topline=true,
bottomline=true,
backgroundcolor=gray!15,
linecolor=gray,
innerleftmargin=5pt,
innerrightmargin=5pt,
innertopmargin=5pt,
innerbottommargin=5pt,
leftmargin=0cm,
rightmargin=0cm,
linewidth=0.75pt]{eBox3}

\definecolor{Red}{RGB}{214, 39, 40}
\definecolor{Blue}{RGB} {31, 119, 180}
\definecolor{Orange}{RGB}{255, 153, 51}
\definecolor{Purple}{RGB}{178, 102, 255}
\definecolor{Green}{RGB}{44, 160, 44}

\definecolor{vio}{RGB}{19, 130, 164}
\definecolor{vioo}{RGB}{89, 2, 155}
\newcommand{\Comment}[1]{{}}
\definecolor{darkblue}{rgb}{0.15,0.35,0.55}
\definecolor{reddish}{rgb}{0.65, 0.2, 0.2}
\definecolor{darkgreen}{RGB}{50,150,0}
\definecolor{greyish}{rgb}{.90,.90,.90}
\definecolor{greyish2}{rgb}{.96,.96,.96}
\definecolor{greyish3}{rgb}{.37,.37,.37}
\definecolor{darkblue2}{rgb}{0.3,0.4,0.9}
\definecolor{Blue3}{RGB}{31, 119, 180}
\usepackage[linktocpage=true]{hyperref}
\hypersetup{
colorlinks=true,
citecolor=darkblue,
linkcolor=reddish,
urlcolor=darkblue,
pdfauthor={},
pdftitle={},
pdfsubject={}
}

\usepackage{colortbl}
\definecolor{lightgreen}{cmyk}{0.2, 0, 0.2, 0.2}
\definecolor{lightgray2}{cmyk}{0.1,0.1,0,0.1}
\definecolor{Red2}{RGB}{214, 39, 40}
\definecolor{Blue2}{RGB} {31, 119, 180}
\definecolor{Orange2}{RGB}{255, 127, 14}
\definecolor{Green2}{RGB}{44, 160, 44}

\makeatletter
\newlength{\apb@width}
\newcommand{\autoparbox}[2][c]{\settowidth{\apb@width}{#2}\parbox[#1]{\apb@width}{#2}}

\makeatother


\def\hs{\hskip 1pt}

\def\beq{\begin{equation}}
\def\eeq{\end{equation}}
\def\be{\begin{equation}}
\def\ee{\end{equation}}

\newcommand{\dif}{\mathrm{d}}

\allowdisplaybreaks[1]
\begin{document}

\newgeometry{top=2cm, bottom=2cm, left=2cm, right=2cm}

\begin{titlepage}
\setcounter{page}{1} \baselineskip=15.5pt 
\thispagestyle{empty}

\begin{center}
{\fontsize{21}{18} \bf Cosmological Collider in the Grassmannian}
\end{center}

\vskip 20pt
\begin{center}
\noindent
{\fontsize{14}{18}\selectfont 
Mattia Arundine\hs$^{1,2}$ and Guilherme L.~Pimentel\hs$^{3}$}
\end{center}

\begin{center}
\vskip8pt
\textit{$^1$ Institute of Physics, University of Amsterdam, Amsterdam, 1098 XH, The Netherlands}

\vskip8pt
\textit{$^2$  Leung Center for Cosmology and Particle Astrophysics,
Taipei 10617, Taiwan}

\vskip 8pt
\textit{$^3$ Scuola Normale Superiore and INFN, Piazza dei Cavalieri 7, 56126, Pisa, Italy}
\end{center}

\vspace{0.4cm}
\begin{center}{\bf Abstract}
\end{center}
\noindent
We revisit the computation of four-point wavefunction coefficients and correlators for external conformally coupled scalars exchanging a particle of generic mass and spin. Much of the phenomenology of cosmological collider physics in the near-de Sitter limit follows from these functions. Computing them in detail is a central challenge in the cosmological bootstrap. Using the cosmological Grassmannian, we write these objects in closed form using hypergeometric functions and Legendre polynomials. We achieve this by writing the standard bootstrap differential equation using the Pl\"ucker coordinates of the Grassmannian, and using the basis of Mandelstam invariants. The exchange in the $s$-channel can be written in terms of a hypergeometric function of the $S$ Mandelstam, while the spin information appears as an overall Legendre polynomial factor that also depends on the other Mandelstams. We fix the boundary conditions by first demanding the absence of unphysical singularities, and, for correlators, by further matching to a kinematic limit in momentum space. Our formulae in Grassmannian space are much simpler than their counterparts in momentum space, demonstrating another useful application of the Grassmannian as a kinematic space for cosmology. 
\end{titlepage}
\restoregeometry

\newpage
\setcounter{tocdepth}{3}
\setcounter{page}{2}

\linespread{1.2}
\tableofcontents
\linespread{1.1}

\newpage

\section{Introduction}

\vskip 4pt
A central challenge that triggered the cosmological bootstrap was the computation of four-point correlators in de Sitter space, particularly those due to the exchange of massive spinning particles. This was a technical hurdle for standard perturbative techniques, but also an important phenomenological benchmark for cosmological collider physics. The resulting expressions in momentum space solve simple differential equations and find several useful applications~\cite{Arkani-Hamed:2015bza, Arkani-Hamed:2018kmz, Jazayeri:2022kjy, Melville:2024ove, Fan:2025scu, Belrhali:2026ktb, Belrhali:2026rkn, Qin:2022fbv, Qin:2023ejc, Maru:2021ezc, Maru:2022bhr}. Although these equations are relatively simple, one may still ask whether a better formulation exists, since the expressions in momentum space are often complicated, in part because they fail to leverage the full kinematical constraints of the de Sitter group.

\vskip 4pt
In this paper, we employ the \textit{orthogonal Grassmannian} $\mathrm{OGr}(4,8)$, which was recently used in \cite{Arundine:2026fbr} to describe theories of massless particles in de Sitter space.\footnote{The orthogonal Grassmannian previously made its appearance in physics, for instance, in the guise of pure spinors in supersymmetric string theories~\cite{Howe:1991bx,Berkovits:2000fe,Berkovits:2001rb}. More recently and more closely related to the setup of cosmology, it appeared in the study of scattering amplitudes in Chern-Simons matter theories~\cite{Lee:2010du,Huang:2012vt,Huang:2013owa,Kim:2014hva,Huang:2014xza}. In that context, it has been applied specifically to trivialize momentum conservation in three-dimensional spinor variables, and also to present an elegant formula for amplitudes of $2n$ particles in terms of cyclic minors.} The objective is to show that both the derivation and the final form of these correlators simplify drastically, when expressed in terms of different kinematic variables. We find a simple formula for both the four-point wavefunction coefficient and the correlator of scalars exchanging a generic single-particle state, as a function of variables that closely resemble the traditional flat-space Mandelstams.

\vskip 4pt
This novel take on the cosmological collider serves two purposes. It uncovers the simplicity of the problem, which was obscured by the traditional momentum-space formulation, further reinforcing the view that the orthogonal Grassmannian is a very convenient kinematic space. Moreover, it broadens the applicability of this new formalism, showing that it can describe correlators with massive internal particles, rather than being restricted to theories with only massless fields.

\subsection*{Strategy and Summary}
The steps that determine the formula for the exchange of a massive particle are straightforward. The setup requires some knowledge of the orthogonal Grassmannian (which we review below) and of the action of differential operators in this space.

\vskip 4pt
We will consider both four-point wavefunction coefficients $\psi_4$, which encode essential information about correlation functions on the future boundary of the four-dimensional de Sitter space dS$_4$, and correlators themselves. In Anti-de Sitter, these are boundary four-point correlators of normalizable modes. The kinematic properties of these objects are identical to those obeyed by correlation functions of primary operators in three-dimensional conformal field theories. Throughout this paper, we will always consider the external particles to be conformally coupled scalars, whose dual operators are scalar primaries with scaling dimension $\Delta = 2$. We will equivalently refer to wavefunction coefficients as \textit{correlators} whenever the distinction is not relevant. Our approach can be generalized to wavefunction coefficients of external massless particles with spin $\ell$, whose dual operators are conserved currents $J^{\mu_1 \dots \mu_\ell}$ with scaling dimension $\Delta = \ell+1$. 

\vskip 4pt
This special class of wavefunction coefficients satisfies several constraints imposed by both conformal symmetry and---in the case of spinning particles---current conservation, which can be automatically accounted for via the following integral representation \cite{Arundine:2026fbr}:
\be
    \label{eq:mainint}
    \psi_4(\Lambda) = \int \dif C \: \delta(C \cdot \Lambda) \: A_4(C) \,,
\ee
where $C$ is a $4 \times 8$ matrix that identifies an element of the orthogonal Grassmannian $\mathrm{OGr}(4,8)$, and $\Lambda$ is an $8 \times 2$ matrix constructed out of the cosmological spinor helicity variables $\{ \lambda_i^\alpha, \bar \lambda_i^\alpha\}$, with $\alpha = 1,2,$ for each particle $i = 1, \dots, 4$. These variables provide a convenient description of the kinematics of wavefunction coefficients, and correlators of external scalars.

\vskip 4pt
This representation shifts the study of $\psi_4$ to the study of the integrand $A_4$, which is subject to useful homogeneity constraints demanded by ``little group''\footnote{The little group is the scaling redundancy in the spinor helicity variables describing a fixed three-momentum. It manifests itself as a scaling property of the determinants appearing in the Grassmannian correlators.} covariance and is expected to be much simpler. We will focus on massive exchanges in the $s$-channel at tree level, which are known to satisfy the following differential equation \cite{Arkani-Hamed:2018kmz, Costa:2011dw, Eberhardt:2020ewh, Herderschee:2022ntr}:
\be
    \label{eq:mastdiffeq}
    \Big[-\Box_{12} + (\Delta(3-\Delta) - J(J+1))\Big] \, \psi_{4,s}^{\Delta,J}(\Lambda) = \psi^J_{4, \rm c}(\Lambda) \,,
\ee
where the right-hand side is a contact solution whose form depends only on the spin $J$ of the exchanged particle, and $\Box_{12}$ is a second-order differential operator acting on particles $1$ and $2$. More precisely, it is the momentum-space representation of the quadratic Casimir of the isometry group $\mathrm{SO}(4,1)$ of dS$_4$. A second equation arises from applying $\Box_{34}$, thus yielding two separate partial differential equations coupled by the same source term. The inability to encode all isometries is ultimately the source of two, rather than one, differential equations, which complicates the solution.

\vskip 4pt
The easiest path to the exchange diagram in the Grassmannian is then to rewrite the above equation in this different kinematic space, where it will take the same form
\be
    \label{eq:mastdiffeqGrass}
    \Big[-\mathcal{C}_{12} + (\Delta(3-\Delta) - J(J+1))\Big] \, A_{4,s}^{\Delta,J}(C) = A^J_{4, \rm c}(C) \,.
\ee
Leveraging the constraints on $A_{4,s}^{\Delta,J}$ will allow us to write it in a convenient form that only depends on the special coordinates $S, T, U$, thus exposing a useful analogy between correlators in the Grassmannian and the associated flat-space amplitudes. This form will also turn the above equation into an ordinary differential equation in \textit{one} variable. In particular, all functions of the form \eqref{eq:mainint} solve the equation that would come from the application of $\mathcal{C}_{34}$. It is therefore unnecessary to impose it separately. Given this differential equation, we follow a similar strategy as in momentum space, finding a particular solution and then imposing suitable boundary conditions.

\vskip 4pt
Our main result is the correlator of four conformally coupled scalars exchanging a particle of spin $J$ and scaling dimension $\Delta$ in the $s$-channel:
\begin{align}
    A_{4,s}^{\Delta,J} & = \:\frac{\mathcal{N}(\Delta,J)}{\mathcal{E}} (2S)^J P_J \bigg( \frac{U-T}{S} \bigg) \Bigg({}_3F_2\Bigg[\begin{array}{c} 1,J+1,J+1\\[2pt] \Delta+J, 3-\Delta+J\end{array}\Bigg| \, 2S \Bigg] \nonumber \\[4pt]
    & + \frac{c_1(\Delta)}{(2S)^{2-\Delta+J}} \, {}_2F_1\Bigg[\begin{array}{c} \Delta-1, \Delta-1\\[2pt] 2\Delta-2 \end{array}\Bigg| \, 2S \Bigg] + \frac{c_2(\Delta)}{(2S)^{\Delta-1+J}} \, {}_2F_1\Bigg[\begin{array}{c} 2-\Delta,2-\Delta\\[2pt] 4-2\Delta \end{array}\Bigg| \, 2S \Bigg]  \Bigg) \,,
\end{align}
where $\mathcal{N}(\Delta,J)$ is a normalization constant and the coefficients $c_{1,2}(\Delta)$ will be determined in the main text. The result splits naturally into a hypergeometric $_3F_2$ function, which contains all the EFT information of the diagram, and the particle production terms proportional to ${}_2F_1$. Once the integrand has been found, the conversion of the result to momentum space can be carried out via the integral \eqref{eq:mainint}, which can be computed explicitly by contour integration.\footnote{A new representation of the scalar exchange in momentum space emerges as a byproduct, whose most prominent features are that it is expressed as a one-dimensional integral of a \textit{homogeneous} solution to \eqref{eq:mastdiffeqGrass}, and that it admits a closed form in terms of \textit{Kampé de Fériet} functions for arbitrary complex kinematics.} We will perform the calculations for a few cases, as a cross-check of our new formulae.

\subsection*{Outline}
The paper is organized as follows. In Section~\ref{sec:review}, we review the kinematics of boundary correlators in de Sitter space and their representation as integrals over the cosmological Grassmannian, which allows us to turn our attention to the Grassmannian integrands. In Section~\ref{sec:Casimir}, we determine the Grassmannian representative of the quadratic Casimir of $\mathrm{SO}(3,2)$ and link it to EFTs. This is an essential ingredient for the differential equation whose solution is the main result of this paper. In Section~\ref{sec:scalarexch}, we study the special case of the exchange of scalars, whose structure we explain in terms of EFT and particle production contributions. In Section~\ref{sec:spinexch}, we extend the analysis to the exchange of any spinning particle. We further exhibit how the general result simplifies to a rational function for massless and partially massless particles. In Section~\ref{sec:conversion}, we explicitly relate the main Grassmannian results to the known formulae in momentum space. Our conclusions are presented in Section~\ref{sec:conclusions}.

\vskip 4pt
Two appendices contain supplemental material. In Appendix~\ref{app:disc}, we prove that the absence of undesired singularities in the Grassmannian representation of the exchanges enforces an important relation between the two integration constants of the differential equations we solve in the main text. Finally, in Appendix~\ref{app:badcomp}, we compute the integral that relates the Grassmannian expression for the exchange of scalars to momentum space, first in a useful kinematic limit and then in full generality. The former, in particular, allows us to finally fix the integration constants for correlators.

\section{Setup and Review}
\label{sec:review}
In this section, we will give a quick overview of the setup and review the necessary background. We will focus on the de Sitter case, but the extension to Anti-de Sitter is straightforward.

\subsection{Particles in de Sitter Space}
The flat slicing of four-dimensional de Sitter is described by the following line element
\be
    \dif s^2 = \frac{L^2}{\eta^2}(-\dif \eta^2 + \dif x^2) \,,
\ee
where $L$ is the de Sitter radius, $-\infty < \eta < 0$ is conformal time, and $x^\mu, \: \mu=1,2,3,$ are the spatial coordinates. We want to describe equal-time correlation functions of fields on the three-dimensional boundary at $\eta \approx 0$. A scalar particle, in particular, is described by a field $\Phi(\eta,x^\mu)$, which behaves close to the boundary as
\be
    \Phi(\eta, x^\mu) \xrightarrow{\ \eta \to 0\ } \bar \phi(x^\mu) \, \eta^{\bar\Delta} + \phi(x^\mu) \, \eta^\Delta \,,
\ee
where $\bar \Delta$ is the scaling dimension of the field and $\Delta \equiv 3-\bar\Delta$. The scaling dimension is related to the mass $M$ of the particle via $M^2 L^2 = \bar\Delta(3-\bar\Delta)$, which implies
\be
    \bar\Delta = \frac{3}{2} - \sqrt{\frac{9}{4}- M^2 L^2} \,,
\ee 
i.e.~$\bar\Delta$ is always taken to be the smallest root (when real). By this choice, the ``growing'' mode on the boundary is always the field $\bar \phi(x^\mu)$ for $M^2 L^2 < 9/4$.

\vskip 4pt
We can construct a wavefunctional $\Psi[\bar \phi]$, which describes the amplitude of the boundary profile in the vacuum state. It admits the following expansion
\be
    \Psi[\bar\phi] \approx \exp \left( -\sum_{n=2}^{+\infty} \frac{1}{n!} \int \dif^3 x_1 \dots \dif^3 x_n \: \psi_n(x_1^\mu, \dots, x_n^\mu) \, \bar \phi(x_1^\mu) \cdots \bar \phi(x_n^\mu)  \right),
\ee
where $\psi_n$ are the ``wavefunction coefficients.'' The isometry group $\mathrm{SO}(4,1)$ of dS$_4$ acts on them as conformal transformations, so they can be interpreted as correlation functions of dual primary operators $\mathcal{O}(x^\mu)$ in a three-dimensional conformal field theory with scaling dimension $\Delta$:
\be
    \psi_n(x_1^\mu, \dots, x_n^\mu) \equiv \langle \mathcal{O}(x^\mu_1) \cdots \mathcal{O}(x^\mu_n) \rangle \,.
\ee
For bulk massless particles of spin $\ell$, the dual operators are currents $J^{\mu_1 \dots \mu_\ell}$ with scaling dimension $\Delta = \ell +1$ that obey the conservation equation $\partial_{\mu_1} J^{\mu_1 \dots \mu_\ell}=0$. Wavefunction coefficients can then be related to the boundary correlation functions $\langle \Phi(0, x_1^\mu) \cdots \Phi(0, x_n^\mu) \rangle$, which also transform like correlators of primary operators with scaling dimension $\bar\Delta$ in a three-dimensional conformal field theory.

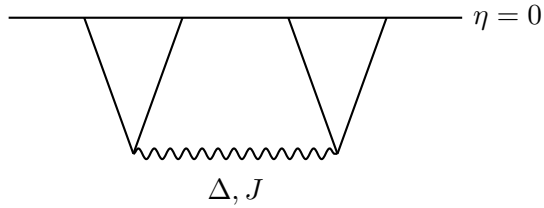
\begin{figure}
\centering
\begin{tikzpicture}[thick, scale=1]
    \draw (-3,0) -- (3,0);
    \node at (3.6,0) {$\eta=0$};

    \coordinate (a1) at (-2,0);
    \coordinate (a2) at (-0.7,0);
    \coordinate (a3) at (0.7,0);
    \coordinate (a4) at (2,0);

    \coordinate (v1) at (-1.35,-1.8);
    \coordinate (v2) at (1.35,-1.8);

    \draw (a1) -- (v1);
    \draw (a2) -- (v1);
    \draw (a3) -- (v2);
    \draw (a4) -- (v2);

    \draw[decorate, decoration={snake, amplitude=2pt, segment length=6pt}] (v1) -- (v2);

    \node at (0,-2.3) {$\Delta, J$};
\end{tikzpicture}
\caption{The Feynman diagram encoding the tree-level contribution to $\langle \mathcal{O} \mathcal{O} \mathcal{O} \mathcal{O} \rangle$ of the exchange of a particle with scaling dimension $\Delta$ and spin $J$. We will always consider external scalars with $\Delta = 2$.}
\label{fig:diagram}
\end{figure}

\vskip 4pt
Both the wavefunction coefficients and the correlators admit a perturbative expansion in terms of Feynman diagrams that depend on the details of the bulk theory. We will focus on the leading contribution to the $n=4$ correlator of a \textit{conformally coupled scalar} $\Phi$ with scaling dimension $\bar\Delta = 1$ due to the interaction with a different field $\Sigma$ with spin $J$ and arbitrary scaling dimension. The relevant Feynman diagram is shown in Figure~\ref{fig:diagram}.

\subsection{Cosmological Grassmannian}
In momentum space, correlators depend on the spatial momenta $k^\mu_i$ and polarization vectors $\epsilon^\mu_i$ of each external particle $i = 1,\dots,4$. Momenta can be equivalently described via the cosmological spinor helicity variables $\lambda_i, \bar \lambda_i$ \cite{Maldacena:2011nz}. We construct a $2 \times 2$ matrix by contracting the three-momentum $k_i^\mu$ with the Pauli matrices ${(\sigma^\mu)_\alpha}^\beta$ (for $\alpha,\beta = 1,2$), and represent the result as an outer product of two-component spinors:
\be
    k_{i,\mu}(\sigma^\mu)_{\alpha \beta} = \lambda_{i,(\alpha} \bar \lambda_{i,\beta)} \,,
\ee
where $(\sigma^\mu)_{\alpha\beta} \equiv \epsilon_{\beta\gamma} {(\sigma^\mu)_\alpha}^\gamma$, with $\epsilon_{\beta\gamma}$ the Levi-Civita symbol. The momenta are invariant under \textit{little group transformations} $\lambda_i \mapsto \rho \lambda_i,  \: \bar \lambda_i \mapsto \rho^{-1} \bar \lambda_i$. Spinor indices can be raised and lowered with the Levi-Civita symbols $\epsilon_{\alpha\beta}$ and $\epsilon^{\alpha\beta}$, allowing us to define the spinor brackets
\be
    \langle i j \rangle \equiv \epsilon_{\alpha\beta} \lambda_i^\alpha \lambda_j^\beta, \quad \langle \bar\imath \bar\jmath \rangle \equiv \epsilon_{\alpha\beta} \bar \lambda_i^\alpha \bar \lambda_j^\beta, \quad \langle i \bar\jmath \rangle \equiv \epsilon_{\alpha\beta} \lambda_i^\alpha \bar \lambda_j^\beta \,.
\ee
In particular, $\langle \bar \imath i \rangle = 2k_i$, with $k_i = |\vec{k}_i|$ the energy of the leg $i$. We can conveniently package these spinors in the following $8 \times 2$ matrix:
\be
    \Lambda \equiv \begin{pmatrix}
        \lambda_1^1 & \lambda_1^2 \\
        \vdots & \vdots \\
        \lambda_4^1 & \lambda_4^2 \\
        \bar \lambda_1^1 & \bar\lambda_1^2 \\
        \vdots & \vdots \\
        \bar \lambda_4^1 & \bar \lambda_4^2
    \end{pmatrix}.
\ee
We take $\Lambda$ to be real, i.e.~$k_i^\mu$ to be spacelike three-momenta in a $(2+1)$-dimensional Lorentzian boundary. All our results can then be related to the Euclidean boundary of dS$_4$ by analytic continuation. It was shown in \cite{Arundine:2026fbr} that the class of wavefunction coefficients of interest admits the following integral representation
\be
    \label{eq:masterint}
    \hat \psi_4(\Lambda) \equiv \left( \prod_{i=1}^4 \frac{1}{k_i^{\Delta_i-2}} \right) \psi_4(\Lambda) = \int \frac{\dif^{4 \times 8} C}{\mathrm{GL}(4)} \, \delta(C \cdot Q \cdot C^T) \, \delta(C \cdot \Lambda) \, A_4(C) \,,
\ee
with $C$ a $4 \times 8$ matrix, and 
\be
    Q \equiv \begin{pmatrix}
        0 & 1_{4 \times 4} \\
        1_{4 \times 4} & 0
    \end{pmatrix}.
\ee 
In the following, we will always work with the rescaled wavefunction coefficients introduced above, and keep the rescaling implicit. The representation \eqref{eq:masterint} can be interpreted as an integral over the \textit{orthogonal Grassmannian} $\mathrm{OGr}(4,8)$, i.e.~the space of null $4$-planes in $\mathbb{R}^{8}$. The $\mathrm{GL}(4)$ quotient is to account for $C$ and $R \cdot C$---with $R \in \mathrm{GL}(4)$---describing the same $4$-plane. We can gauge fix this redundancy by considering the following form of the matrix $C$:
\be
    \label{eq:rightchart}
    C = \begin{pmatrix}
        1_{4 \times 4}, & C_4
    \end{pmatrix},
\ee
with $C_4$ a skew-symmetric $4 \times 4$ matrix, with six independent entries $(C_4)_{ij} = -c_{ij}$, which naturally satisfies the constraint $C \cdot Q \cdot C^T = 0$. Given this quadratic constraint for $C$, the orthogonal Grassmannian decomposes into two disconnected components, which we refer to as ``left'' and ``right'' branches. These need to be gauge-fixed individually, but all wavefunction coefficients of external scalars localize in the branch charted by \eqref{eq:rightchart}. We stress that any wavefunction coefficient that obeys the de Sitter isometries and current conservation admits the form \eqref{eq:masterint}.\footnote{In the case of scalars, the requirement of current conservation is replaced by fixing their mass to be the conformally coupled value $M^2L^2=2$. External massless spinning particles can also be considered~\cite{Arundine:2026fbr}, but we will not do so in this paper.} Correlation functions of external scalars obey the same isometries, hence we can represent them in the Grassmannian in the same way.

\begin{figure}[t!]
	\centering
	\includegraphics[scale=1.19]{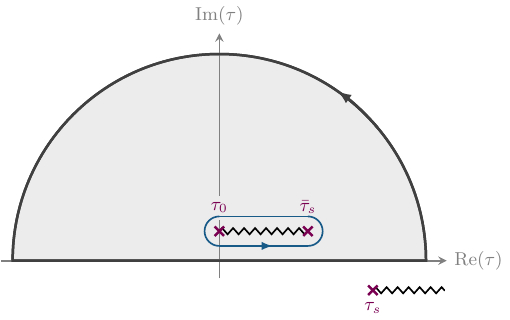} 
	\caption{Integration contour for \eqref{eq:simpleint} (in black) and analytic structure of generic $A_{4,s}$, which follows from a branch cut for $S<0$. The contour will always be deformed to the one in blue unless explicitly stated. The behavior of $A_{4,s}$ on the real line has been regulated via the prescription $\tilde S \mapsto \tilde S+i \varepsilon$ for each Mandelstam. We recall that all spinors are real, hence $\bar\tau_s$ and $\tau_s$ are real as well. For special values of $\Delta$, the branch points become poles and the integral reduces to a sum over the residues in $\tau = 0$ and $\tau = \bar \tau_s$.}
	\label{fig:contour}
\end{figure}

\vskip 4pt
The function $A_4(C)$ only depends on $C$ through its minors
\be
    (I_1 \cdots I_4) = \epsilon^{a_1 \dots a_4} C_{a_1 I_1} \cdots C_{a_4 I_4} \,,
\ee
with $\epsilon^{a_1 \dots a_4}$ the Levi-Civita symbol. Each external particle with spin $\ell_i$ has helicity $h_i = \pm \ell_i$. Labeling the columns of $C$ as $I = \bar 1, \dots, \bar 4, 1, \dots,4,$ the information about the external helicities is encoded as follows. Under little group transformations, which are realized on $C$ via
\be
    C \mapsto C \cdot \rho^{-1}, \quad \rho \equiv \mathrm{diag} \left(\rho_1, \dots, \rho_4, \frac{1}{\rho_1}, \dots, \frac{1}{\rho_4} \right),
\ee
the correlator transforms as
\be
    \label{eq:A4constraint1}
    A_4((I_1 \cdots I_4)) \mapsto \left( \prod_{i=1}^4 \rho_i^{-2h_i} \right) A_4((I_1 \cdots I_4)) \,.
\ee
Furthermore, in order for the integral \eqref{eq:masterint} to be well-defined projectively, under a $\mathrm{GL}(4)$ transformation $C \mapsto R \cdot C$, it must hold that
\be
    \label{eq:A4constraint2}
    A_4((I_1 \cdots I_4)) \mapsto \det(R)^{-1} A_4((I_1 \cdots I_4)) \,.
\ee
Integrating out the delta functions in \eqref{eq:masterint} in the gauge \eqref{eq:rightchart} simplifies the expression to a one-dimensional integral
\be
    \label{eq:simpleint}
    \psi_4 = \delta(\vec k_1 + \vec k_2 + \vec k_3 + \vec k_4)  \: \mathcal{J} \int \frac{\dif \tau}{2 \pi i} A_4(c_{ij}(\tau)) \,,
\ee
where $\mathcal{J}$ is a numerical Jacobian factor that we will ignore, and the six parameters $c_{ij}$ evaluate to
\be
    c_{ij}(\tau) = \frac{\langle i j\rangle}{E} + \tau \frac{\epsilon_{ijkl} \langle \bar k \bar l \rangle}{2} \,,
\ee
with $E = k_1+k_2+k_3+k_4$. The contour of integration for $\tau$ is shown in Figure~\ref{fig:contour}, and maps the $A_4$ found in this paper to the desired $\psi_4$.

\subsubsection*{Mandelstam variables}
It turns out that any little group-invariant function can be expressed in terms of the following ``Mandelstam variables''
\be
    \tilde S \equiv (\bar 1 \bar 2 1 2), \quad \tilde T \equiv (\bar 1 \bar 4 1 4), \quad \tilde U \equiv (\bar 1 \bar 3 1 3) \,.
\ee
Expressed in terms of $\tau$, they take a very simple form:
\be
    \tilde S(\tau) = \langle \bar 1 \bar 2 \rangle \langle \bar 3 \bar 4 \rangle (\tau-\tau_s)(\tau-\bar\tau_s) \,,
\ee
where we have defined
\be
    \tau_s \equiv \frac{1}{\langle \bar 1 \bar 2 \rangle \langle \bar 3 \bar 4 \rangle} \frac{E_L E_R}{E}\,, \quad \bar \tau_s \equiv \frac{1}{\langle \bar 1 \bar 2 \rangle \langle \bar 3 \bar 4 \rangle} \frac{\bar E_L \bar E_R}{E} \,,
\ee
with $k_s \equiv |\vec k_1 + \vec k_2|$ and
\be
\begin{aligned}
    \label{eq:somesings}
    E_L&\equiv k_1+k_2+k_s\,, \qquad \bar E_L \equiv k_1+k_2-k_s\,,\\
    E_R&\equiv k_3+k_4+k_s\,, \qquad \bar E_R \equiv k_3+k_4-k_s\,.
\end{aligned}
\ee
The sum of these Mandelstams is not zero, but rather proportional to the total energy $E$
\be 
{\cal E}(\tau)\equiv -\frac{(\tilde S+ \tilde T+ \tilde U)(\tau)}{2}= \tau\hs E\,.
\ee
For external scalars, $A_4$ is little group-invariant and obeys \eqref{eq:A4constraint2}. By defining the $\mathrm{GL}(4)$-invariant Mandelstams
\be
    S \equiv \frac{\tilde S}{\tilde S + \tilde T + \tilde U} \,, \quad T \equiv \frac{\tilde T}{\tilde S + \tilde T + \tilde U} \,, \quad U \equiv \frac{\tilde U}{\tilde S + \tilde T + \tilde U} \,,
\ee
the most general Grassmannian integrand takes the following form
\be
    A_4 = \frac{1}{\mathcal{E}} F(S,T,U) \,,
\ee
where the three rescaled Mandelstams satisfy $S+T+U = 1$. In terms of two independent variables, the study of $s$-channel diagrams will become particularly simple by writing
\be
    \label{eq:generalscalar}
    A_{4,s} = \frac{1}{\mathcal{E}} F\bigg(S, \frac{U-T}{S} \bigg) \,.
\ee
The exchanges in the other channels follow from trivial permutations of the external particles.\footnote{For external particles with spin $\ell_i$, we must satisfy \eqref{eq:A4constraint1}. This is easily accommodated. For currents with identical spin $\ell_i = \ell$ and e.g. helicities $\{ -,-,+,+ \}$, the uplift is simply
\begin{equation*}
    A_{4,s}^{--++} = \frac{(1 2 \bar 3\bar 4)^{2\ell}}{\mathcal{E}^{2\ell+1}} F\bigg(S, \frac{U-T}{S} \bigg) \,.
\end{equation*}
We will not further pursue these examples, but the analysis presented here applies to them in the same way.}

\section{Casimir Operator, EFTs and Exchanges}
\label{sec:Casimir}
As mentioned in the introduction, the desired correlators are solutions of a differential equation. Since this equation features the quadratic Casimir of the isometry group $\mathrm{SO}(4,1)$ of dS$_4$, we now discuss its properties and derive how it acts on the function $A_4(C)$ appearing in \eqref{eq:masterint}.

\subsection{A Lesson from Flat Space}
\label{ssec:lessonflat}
As a warm-up, we consider $2 \to 2$ scattering in flat space. Mandelstam variables can be characterized as the quadratic Casimir of the Poincaré group for pairs of particles. For example, we obtain the variable $s$ by considering particles $1$ and $2$ and the corresponding generators of translations, $P_1^\mu$ and $P_2^\mu$: 
\be
    s \equiv -(P_1^\mu + P_2^\mu)^2 = \mathcal{C}_{12}^{\rm flat} \,.
\ee
It is then natural, in the case of dS$_4$, to think of the quadratic Casimir of $\mathrm{SO}(4,1)$ instead. Finding a representation of this Casimir in the Grassmannian is most easily achieved via twistors and a suitable analytic continuation from the Casimir of $\mathrm{SO}(3,2) \simeq \mathrm{Sp}(4,\mathbb{R})$, i.e.~the symplectic group.

\vskip 4pt
The quadratic Casimir has interesting properties. Consider, for example, the wavefunction coefficient associated with four conformally coupled scalars with a $\Phi^4$ coupling:
\be
    \langle \mathcal{O}(k_1) \mathcal{O}(k_2) \mathcal{O}(k_3) \mathcal{O}(k_4) \rangle' = \frac{1}{E} \,,
\ee
where the prime denotes that we omit the three-momentum conserving delta function. The repeated application of the Casimir on the function above generates the EFT interactions arising from the exchange of a scalar $\Sigma$ with mass $M$ that interacts with the field $\Phi$ via $\Phi^2 \Sigma$ \cite{Green:2024cmx}. Indeed, we find e.g.~in the $s$-channel that\footnote{Strictly speaking, this is not an expansion in any small parameter, as we expand in powers of $\Box_{12}/M^2$.}
\be
	\langle \mathcal{O}(k_1) \mathcal{O}(k_2) \mathcal{O}(k_3) \mathcal{O}(k_4) \rangle' \propto \frac{1}{-\Box_{12}+M^2} \Bigg[ \frac{1}{E} \Bigg] \approx \frac{1}{M^2} \sum_{n=0}^{+\infty} \frac{\Box_{12}^n}{M^{2n}} \Bigg[ \frac{1}{E} \Bigg] \,,
\ee
where the box operator is exactly the Casimir of $\mathrm{SO}(4,1)$. The above expansion misses non-perturbative terms in the mass due to spontaneous particle production. This perfectly matches the form of the corresponding flat-space amplitude, where momentum space naturally diagonalizes the action of $\mathcal{C}_{12}^{\rm flat}$, thus making its action \textit{algebraic} rather than differential:
\begin{equation}
	\mathcal{A}_s = \frac{1}{-s+M^2} = \frac{1}{M^2} \sum_{n=0}^{+\infty} \frac{s^n}{M^{2n}} [1] \,.
\end{equation}
In the Grassmannian, the Casimir operator will remain a differential operator, but its action will be much simpler than in momentum space.

\subsection{Casimir from (Dual) Twistors}
We first find the quadratic Casimir in (dual) twistor space, restricting ourselves to external conserved currents or conformally coupled scalars. One can think of (dual) twistors as the four-component (anti-)spinorial representations $Z^A \hs (W_A)$ of the symplectic group $\mathrm{Sp}(4,\mathbb{R}) \simeq \mathrm{SO}(3,2)$, whose index can be lowered and raised via the symplectic form
\be
    \Omega_{AB} = \begin{pmatrix}
			0 & 1_{2\times 2} \\
			- 1_{2 \times 2} & 0
		\end{pmatrix}, \quad \Omega^{AB} = -\Omega_{AB} \,.
\ee
For example, $W^A \equiv \Omega^{AB} W_B$ and $W_A \equiv \Omega_{AB} W^B$, with contractions defined as 
\be
\begin{aligned}
    W_1 \cdot W_2 & \equiv W_{1,A} \Omega^{AB} W_{2,B}\,,\\
    W_1 \cdot \partial_{W_2} & \equiv W_{1,A} \partial_{W_{2,A}} \,, \\
    \partial_{W_1} \cdot \partial_{W_2} & \equiv -\partial_{W_{1,A}} \Omega_{AB} \partial_{W_{2,B}} \,.
\end{aligned}
\ee
In the following, we will focus our attention on dual twistors for simplicity. A dual twistor can be assigned to each conserved current, and this choice furnishes another remarkable kinematic space for wavefunction coefficients of conserved currents \cite{Baumann:2024ttn,CarrilloGonzalez:2025qjk}. This is conceptually similar to the use of spinor helicity variables in flat space, where the spinors of the Lorentz group $\mathrm{SO}(3,1)$ can be used to describe scattering amplitudes of massless particles.

\vskip 4pt
An element of the isometry group $\mathrm{Sp}(4,\mathbb{R})$ acts on dual twistors via a simple ``rotation''
\begin{equation}
	F(W_A) \to F({M_A}^B W_B), \quad M \in \mathrm{Sp}(4,\mathbb{R}) \,.
\end{equation}
An infinitesimal transformation ${M_A}^B = {\delta_A}^B + {X_A}^B$, with $X^{AB} = X^{BA}$, therefore yields
\begin{align}
	F(W_A + {X_A}^B W_B) & \simeq F(W_A) + {X_C}^B W_B \frac{\partial}{\partial W_C} F(W_A) \nonumber \\[4pt]
		& \simeq F(W_A) - X^{DB} W_B \Omega_{DC} \frac{\partial}{\partial W_C} F(W_A) \nonumber \\[4pt]
		& \simeq F(W_A) + \frac{1}{2} X^{BD} J_{BD} [F(W_A)] \,,
\end{align}
from which we can read the following ten differential operators:
\begin{equation}
	J_{AB} = J_{BA} =-W_A \Omega_{BC} \frac{\partial}{\partial W_C} - W_B \Omega_{AC} \frac{\partial}{\partial W_C} \equiv -W_A \frac{\partial}{\partial W^B} - W_B \frac{\partial}{\partial W^A} \,.
\end{equation}
It is straightforward to determine their commutation relations:
\begin{equation}
	[J_{AB},J_{CD}] = \Omega_{CB} J_{AD} + \Omega_{CA} J_{BD} + \Omega_{DB} J_{AC} + \Omega_{DA} J_{BC} \,.
\end{equation}
These commutators are consistent with the fact that the operators $J_{AB}$ realize a representation of the algebra of $sp(4,\mathbb{R}) \simeq so(3,2)$ on functions of dual twistors (i.e., for our purposes, correlators). These generators can then be used to compute the quadratic Casimir:
\be
	\mathcal{C}_{12} = \frac{1}{4} (J_{1,AB} + J_{2,AB}) (J_1^{AB} + J_2^{AB}) \,.
\ee
A careful computation\footnote{In our conventions, $\partial_{W_A} W_B = \delta^A_B$, but $\partial_{W^B} W^A = -\delta^A_B$.} shows that the Casimir takes the following form:
\begin{align}
    \mathcal{C}_{12} = & -2(\ell_1^2 + \ell_2^2 - 2) + (W_1 \cdot W_2)(\partial_{W_1} \cdot \partial_{W_2}) + h_1 + h_2 - 2 \nonumber \\[4pt]
    & - \frac{1}{2} (W_1 \cdot\partial_{W_2}) (W_2 \cdot \partial_{W_1}) - \frac{1}{2} (W_2 \cdot\partial_{W_1}) (W_1 \cdot \partial_{W_2}) \,,
    \label{eq:twistorcas}
\end{align}
where $\ell_i$ are the spins of the external particles, and $h_i$ are their helicities. We have used that $F(\rho W_i) = \rho^{2h_i-2} F(W_i)$, which is demanded by little group-covariance of the correlators in twistor space.

\vskip 4pt
We have based our presentation on the dual twistors $W_{i,A}$. In terms of the twistors $Z^A$, the generators simply turn out to be
\begin{equation}
	J_{AB} = Z_A \frac{\partial}{\partial Z^B} + Z_B\frac{\partial}{\partial Z^A} \,,
\end{equation}
for which the Casimir can be determined in a similar fashion.

\subsection{Acting in the Grassmannian}
The orthogonal Grassmannian arises from (dual) twistor space via a Schwinger parametrization of the $\mathrm{Sp}(4,\mathbb{R})$-invariant correlators, which must be constructed from inner products of (dual) twistors
\be
    F(W_i \cdot W_j) = \int \dif^6 c_{ij} \: A_4(c_{ij}) \, \exp\left( -\frac{i}{2} c_{ij} W_i \cdot W_j \right) .
\ee
The action of the Casimir in the Grassmannian can then be found through differentiation and integration by parts, starting from \eqref{eq:twistorcas}. The Schwinger parametrization above describes exactly the chart \eqref{eq:rightchart} on the right branch. One can then show that the following expression holds:
\begin{equation}
	\begin{aligned}
    \label{eq:terribleCasimir}
	\mathcal{C}_{12} [A_4] = & -5 \partial_{c_{12}} (c_{12} A_4) - c_{14} c_{23} \partial_{c_{24}} \partial_{c_{13}} A_4 - c_{13} c_{24} \partial_{c_{14}} \partial_{c_{23}} A_4 \\[4pt]
	& -\partial_{c_{13}} \partial_{c_{23}} (c_{13} c_{23} A_4) - \partial_{c_{14}} \partial_{c_{24}} (c_{14} c_{24} A_4) + \sum_{(ij) \neq (34)} \partial_{c_{12}} \partial_{c_{ij}} (c_{12} c_{ij} A_4) \\[4pt]
	& + (c_{13} c_{24} - c_{23} c_{14}) \partial_{c_{12}} \partial_{c_{34}} A_4 - 2(\ell_1^2 + \ell_2^2 - 2) A_4 \,.
	\end{aligned}
\end{equation}
Although this expression is not particularly insightful, we will soon see how it greatly simplifies for physical correlators of the form \eqref{eq:generalscalar}.

\vskip 4pt
As a first application, we check how the Casimir acts on the contact $\Phi^4$ interaction. In the Grassmannian, this is simply described by $A_{4, \rm c} = 1/\mathcal{E}$, as is clear from encircling the $\tau = 0$ pole in \eqref{eq:simpleint}. As illustrated in Section~\ref{ssec:lessonflat}, multiple applications of the Casimir describe the EFT obtained by integrating out a heavy scalar $\Sigma$ with a $\Phi^2 \Sigma$ interaction: 
\begin{equation}
	\begin{aligned}
    \label{eq:firstEFTterms}
	\mathcal{C}_{12} \left[\frac{1}{\mathcal{E}}\right] & = \frac{2(S+1)}{\mathcal{E}} \,, \\[4pt]
	\mathcal{C}_{12}^2 \left[\frac{1}{\mathcal{E}}\right] & = \frac{4(1+S+4 S^2)}{\mathcal{E}} \,, \\[4pt]
     \mathcal{C}_{12}^3 \left[ \frac{1}{\mathcal{E}} \right] & = \frac{8(1+S-4S^2+36S^3)}{\mathcal{E}} \,, \quad \dots \\[4pt]
	\end{aligned}
\end{equation}
The most interesting feature of the Casimir operator is its action on single-channel contributions to the correlators. Indeed, the general structure of the Casimir equation \eqref{eq:mastdiffeq} is naturally formulated in the Grassmannian as
\be
    \label{eq:mastdiffeq2}
    \Big[-\mathcal{C}_{12} + (\Delta(3-\Delta) - J(J+1))\Big] \, A_{4,s}^{\Delta,J}(C) = A^J_{4, \rm c}(C) \,.
\ee
Such a structure parallels the flat-space result
\begin{equation}
	\Big[-s+M^2\Big] \, \mathcal{A}_{s} = \mathcal{A}_{\rm c} \,,
\end{equation}
where $M^2$ is the Casimir of the exchanged particle, $\mathcal{A}_s$ is the exchange diagram in the $s$-channel and $\mathcal{A}_{\rm c}$ is a contact amplitude. In the Grassmannian, we naturally describe \textit{external} particles whose Casimir equals $-2(\ell^2-1)$, i.e.~conserved currents and conformally coupled scalars, but we can still allow for internal masses. In the following, we will solve \eqref{eq:mastdiffeq2} for external scalars, first in the case of the exchange of a scalar of arbitrary scaling dimension $\Delta$, and then for the exchange of a spinning particle. Wavefunction coefficients and correlators will differ by homogeneous terms.

\section{Exchanging Scalars}
\label{sec:scalarexch}
Recall that, for external scalars, the most general $A_4$ admits the simple form
\be
    A_4 = \frac{1}{\mathcal{E}} F\bigg(S, \frac{U-T}{S} \bigg) \,.
\ee
Indeed, it is easy to show that $S, T, U$ furnish a basis for all the combinations of the entries of $C$ that transform trivially under both little group transformations and $\mathrm{GL}(4)$ transformations. The factor of $1/\mathcal{E}$ is then a convenient choice that complies with \eqref{eq:A4constraint2}.

\vskip 4pt
For $s$-channel exchanges of \textit{scalar particles}, a natural ansatz is
\begin{equation}
    \label{eq:scalaransatzz}
    A_{4,s}^\Delta = \frac{1}{\mathcal{E}} F(S) \,.
\end{equation}
The intuition arises from the fact that, for amplitudes, a dependence on $t$ and $u$ for $s$-channel diagrams only arises when exchanging spinning particles. Furthermore, the EFT expansion of the exchange only depends on $S$, as is clear from \eqref{eq:firstEFTterms}. When collapsing the exchange of a scalar particle \textit{with no derivative interactions}, the resulting contact interaction is $\Phi^4$, which we recall is given in the Grassmannian by $A_{4, \rm c} = 1/\mathcal{E}$.\footnote{Exchanges from higher-derivative cubic vertices can always be written as an exchange coming from a cubic operator with lowest mass dimension, plus contact terms.} These observations reduce \eqref{eq:mastdiffeq2} to
\begin{equation}
    \label{eq:firstdiffeq}
    \Big[-\mathcal{C}_{12} + \Delta(3-\Delta)\Big] \left[ \frac{1}{\mathcal{E}} F(S) \right] = \frac{1}{\mathcal{E}} \,,
\end{equation}
where $\Delta = 3/2+\sqrt{9/4-M^2 L^2}$. The action of \eqref{eq:terribleCasimir} simplifies to
\be
    \mathcal{C}_{12}\left[ \frac{1}{\mathcal{E}} F(S) \right] = \frac{1}{\mathcal{E}} \Big[ S^2(2S-1)F''(S) + 2S(3S-1) F'(S) + 2(S+1) F(S) \Big] \,. 
\ee
One must consider the cases $\Delta \in \mathbb{Z}, \: \Delta \in \mathbb{Z}+1/2$ and $\Delta \notin \{ \mathbb{Z}, \mathbb{Z}+1/2 \}$ separately. In the latter case, \eqref{eq:firstdiffeq} is solved by
\begin{align}
    \label{eq:firstsol}
    F(S) = \:\frac{1}{(\Delta-1)(\bar\Delta-1)} \Bigg( & {}_3F_2\Bigg[\begin{array}{c} 1,1,1\\[2pt] \Delta, \bar\Delta\end{array}\Bigg| \, 2S \Bigg] + \frac{c_1(\Delta)}{(2S)^{2-\Delta}} {}_2F_1\Bigg[\begin{array}{c}\Delta-1, \Delta-1\\[2pt] 2\Delta-2 \end{array}\Bigg| \, 2S \Bigg] \nonumber \\[4pt]
    & + \frac{c_2(\Delta)}{(2S)^{2-\bar\Delta}} {}_2F_1\Bigg[\begin{array}{c} \bar\Delta-1, \bar\Delta-1\\[2pt] 2\bar\Delta-2 \end{array}\Bigg| \, 2S \Bigg] \Bigg) \,,
\end{align}
where $c_{1,2}(\Delta)$ are integration constants, and we recall $\bar\Delta \equiv 3-\Delta$. This gives a closed-form expression for the diagram. Furthermore, we will see that each term has a simple physical interpretation. We will fix $c_{1,2}(\Delta)$ later, and for now focus on the particular solution.

\subsection{EFT Expansion}
As a first example, we want to specialize the general answer to the exchange of a \textit{principal series scalar} with $\Delta = \frac{3}{2} + i\lambda, \: \lambda \in \mathbb{R}$. In the EFT limit, $\lambda \to \infty$, the expansion of \eqref{eq:firstsol} for generic values of $S$ is blind to the terms proportional to $c_{1,2}(\Delta)$, so we expect them to be the particle production contributions to the correlator. The choice $c_{1,2}(\Delta) = 0$ is also the only regular solution in the $S \to 0$ limit.

\vskip 4pt
Regardless of the values of $c_1$ and $c_2$, then, the EFT expansion can be read from expanding the particular solution in powers of $S$:
\be
    \label{eq:series3F2}
    {}_3F_2\Bigg[\begin{array}{c} 1,1,1\\[2pt] \Delta, \bar\Delta\end{array}\Bigg| \, 2S \Bigg] = \sum_{n=0}^{+\infty} \frac{(n!)^2}{(\Delta)_n (\bar \Delta)_n} (2S)^n \,,
\ee
where we have used the Pochhammer symbol
\be
    (a)_n \equiv a(a+1)\cdots(a+n-1) = \frac{\Gamma(a+n)}{\Gamma(a)} \,.
\ee
This result features a $\Delta$-dependent coefficient in front of each power of $S$, which is the result of resumming subleading contributions in $1/M^2 = 1/(\Delta(3-\Delta))$ to each coefficient. Note that, unlike the usual EFT expansion of flat-space amplitudes, which converges for $s/m^2 < 1$, the series representation of the ${}_3F_2$ \eqref{eq:series3F2} has a convergence condition on $S$ rather than $S/|\Delta|^2$, i.e.~$|S| < 1/2$. This is because the Pochhammer symbols also grow factorially, so that the series coefficients have a $1/n$ asymptotic behavior, with a logarithmic singularity at $2S=1$.

\subsection{Particle Production}
\label{ssec:partprod}
We finally want to fix $c_1$ and $c_2$ using suitable boundary conditions. First, we impose the absence of unphysical singularities. The generalized hypergeometric function we have found is singular at $S = 1/2$ and has a branch cut for $S > 1/2$. This singularity is the avatar of folded singularities in momentum space,\footnote{A folded singularity can be set to $0$ with a choice of Euclidean boundary momenta, for instance $\bar E_L$ and $\bar E_R$ in~\eqref{eq:somesings}. In other words, it is a singularity in the physical region. These are always absent in correlators and wavefunction coefficients.} since $1-2S = U+T-S$ is expected to be a regular point of our solution. Only $S$ is allowed to be singular, as the exchange examples in \cite{Arundine:2026fbr} show. Setting the discontinuity to $0$ for $S > 1/2$ implies
\be
\begin{aligned}
    \label{eq:generalc1Del}
    c_1(\Delta) & = -\frac{\Gamma(\Delta)\Gamma(\bar\Delta)\Gamma(\Delta-1)^2}{2\Gamma(2\Delta-2)} \Big[ 1 + \tilde f(\Delta) \Big] \,, \\[4pt]
    c_2(\Delta) & = -\frac{\Gamma(\Delta)\Gamma(\bar\Delta)\Gamma(\bar\Delta-1)^2}{2\Gamma(2\bar\Delta-2)} \Big[ 1 - \tilde f(\Delta) \Big] \,,
\end{aligned}
\ee
where $\tilde f(\Delta)$ is a function which distinguishes the correlator from the wavefunction coefficient. We prove this result in Appendix~\ref{app:disc}.
\begin{itemize}
    \item The wavefunction coefficient is most easily determined as the corresponding correlator in AdS$_4$. Indeed, their equality follows easily from a Wick rotation in the bulk integral representation~\cite{Bzowski:2023nef}. In AdS$_4$, the second boundary condition is $c_2(\Delta) = 0$, i.e.~only the normalizable mode with $\Delta_+ \equiv \Delta = 3/2+\sqrt{9/4+M^2L^2}$ propagates in the bulk. This choice implies $\tilde f(\Delta) = 1$.

    \item The correlator can be determined in two ways. In Appendix~\ref{app:badcomp}, we study the $k_s \to 0$ limit of the Grassmannian integrand when converted to momentum space. Demanding that the known \textit{collapsed limit} of the exchange \cite{Arkani-Hamed:2015bza, Arkani-Hamed:2018kmz} is recovered yields
    \be
    \label{eq:scalarftilde}
    \tilde f(\Delta) = \frac{1}{\sin(\pi(\Delta-3/2))} \,.
    \ee
    Alternatively, one can apply the analytic continuation derived in~\cite{Sleight:2021plv} to the AdS$_4$ correlator (see Equation~(4.32)), which returns the same result. We observe that $c_2(\Delta) = c_1(\bar\Delta)$, i.e.~$A_{4,s}^\Delta$ is invariant under $\Delta \leftrightarrow \bar \Delta$ and is thus shadow-symmetric, as expected.
\end{itemize}
This choice of coefficients has two consequences on the analytic structure of our solution: it removes the branch cut for $S > 1/2$, but introduces a branch cut for $S < 0$ (recall that $\Delta$ is non-integer) via the power-law pieces of the homogeneous solutions. This observation will be crucial for the conversion of our expressions to momentum space.

\subsection{Full Solution}
The above analysis has led us to the solution:
\begin{align}
    \label{eq:arbitraryscalar}
    {}&
    \boxed{\begin{aligned}
    A_{4,s}^\Delta = \frac{1}{(\Delta-1)(\bar\Delta-1)} \frac{1}{\mathcal{E}} \Bigg( & {}_3F_2\Bigg[\begin{array}{c} 1,1,1\\[2pt] \Delta, \bar\Delta\end{array}\Bigg| \, 2S \Bigg] + \frac{c_1(\Delta)}{(2S)^{2-\Delta}} \, {}_2F_1\Bigg[\begin{array}{c}\Delta-1, \Delta-1\\[2pt] 2\Delta-2 \end{array}\Bigg| \, 2S \Bigg] \\[4pt]
    & + \frac{c_2(\Delta)}{(2S)^{2-\bar\Delta}} \, {}_2F_1\Bigg[\begin{array}{c} \bar\Delta-1, \bar\Delta-1\\[2pt] 2\bar\Delta-2 \end{array}\Bigg| \, 2S \Bigg] \Bigg) \,,
    \end{aligned}}
\end{align}
where the coefficients $c_{1,2}(\Delta)$ are
\be
\begin{alignedat}{2}
    \label{eq:scalarc1c2dsAds}
    c_1^{(\rm AdS)}(\Delta) & = c_1^{(\rm WF)}(\Delta) && = -\frac{\Gamma(\Delta)\Gamma(\bar\Delta)\Gamma(\Delta-1)^2}{\Gamma(2\Delta-2)} \,, \\[4pt]
    c_2^{(\rm AdS)}(\Delta) & = c_2^{(\rm WF)}(\Delta) && = 0\,, \\[4pt]
    c_1^{(\rm dS)}(\Delta) & = c_2^{(\rm dS)}(\bar\Delta) && = -\frac{\Gamma(\Delta)\Gamma(\bar\Delta)\Gamma(\Delta-1)^2}{2\Gamma(2\Delta-2)} \bigg[ 1 + \frac{1}{\sin(\pi(\Delta-3/2))} \bigg] \,.
\end{alignedat}
\ee
In AdS$_4$, $A_{4,s}^\Delta$ also acquires an overall minus sign due to a sign flip in the left-hand side of \eqref{eq:mastdiffeqGrass}. This follows from the analytic continuation $L_{\rm dS}^2 \mapsto -L_{\rm AdS}^2$. In AdS$_4$, both boundary conditions needed to fix the result can be imposed in the Grassmannian, i.e.~absence of folded singularities and a bulk quantization of the fields with only $\Delta_+$. The collapsed limit then serves as a cross-check {\it a posteriori}. There is no freedom in the choice of $\tilde f(\Delta)$, compatibly with the uniqueness of the object in AdS$_4$.

\vskip 4pt
Our construction works for all $\Delta \notin \{ \mathbb{Z}, \mathbb{Z} + 1/2\}$. In the next subsection, we will see that integer values of $\Delta$ also feature a branch cut for $S < 0$, thus allowing us to define a \textit{unique} contour for the $\tau$ variable when converting these expressions to momentum space. We anticipate that, for generic $\Delta$, the branch cut is due to the homogeneous terms, while for integer $\Delta$ it arises from the particular solution. We will not analyze the half-integer $\Delta$ in this paper.

\subsection{Integer Scaling Dimension}
Let us now determine the wavefunction coefficients $A_{4,s}^\Delta$ for integer $\Delta$. We focus on the cases $\Delta = 2$ and $\Delta = 3$ in dS$_4$, i.e.~internal conformally coupled and massless scalars, respectively. The analysis can be easily extended to other integers.

\vskip 4pt
For $\Delta = 2$, the differential equation~\eqref{eq:firstdiffeq} has the solution
\be
    F(S) = \frac{\mathrm{Li}_2(1-2S) + c_1 + c_2 \log(1-2S)}{2 S} \,.
\ee
Regularity at $S = 1/2$ implies $c_2 = 0$. To select wavefunction coefficients, we impose a softer behavior in the $S \to 0$ limit, which sets $c_1 = -\mathrm{Li}_2(1) = -\pi^2/6$ and yields a logarithmic singularity. This choice is equivalent to demanding $\tilde f(\Delta) = 1$ in \eqref{eq:arbitraryscalar}. The above solution then reduces to
\be
    \label{eq:scalarD2}
    F(S) = \frac{\mathrm{Li}_2(1-2S) - \pi^2/6}{2 S} \,.
\ee
We will verify in Section~\ref{ssec:scalarcheck} that this function reproduces the correct momentum-space object.

\vskip 4pt
For $\Delta = 3$, on the other hand, the differential equation~\eqref{eq:firstdiffeq} has the solution
\begin{align}
    F(S) & = \frac{(S-1) \,\mathrm{Li}_2(1-2S)- (S-1)(2-\log 2)\log(1-2S) + 2S \log S + 2}{2S^2} \nonumber \\[4pt]
    & + c_1 \frac{S-1}{2S^2} + c_2 \frac{2-(S-1) \log(1-2S)}{2S^2} \,.
\end{align}
Regularity at $S = 1/2$ implies $c_2 = -2+\log 2$, while a soft $S \to 0$ behavior implies $c_1 = -\pi^2/6 -2 + \log 4$. The above solution then reduces to\footnote{Despite appearances, the function is regular at $S=0$.} 
\be
    \label{eq:scalarD3}
    F(S) = \frac{\mathrm{Li}_2(1-2S) - \pi^2/6}{2 S} \left( 1 - \frac{1}{S} \right) + \frac{\log 2S - 1}{S} \,,
\ee
where we can clearly see the imprint of the $\Delta = 2$ answer, which can be used as a seed to generate this exchange. The (poly)logarithms are responsible for a branch cut for $S < 0$. One can check that this feature persists for all other integer values of $\Delta$. These formulae can also be obtained from~\eqref{eq:arbitraryscalar} by taking the \textit{finite} $\Delta \to 2$ and $\Delta \to 3$ limits, respectively. Correlators can be easily extracted in the same way.

\section{Exchanging Spinning Particles}
\label{sec:spinexch}
In this section, we show how to describe conformally coupled scalars exchanging particles of spin $J$ and \textit{generic} values of $\Delta$ in the Grassmannian. In this case, a natural ansatz for the correlator \eqref{eq:generalscalar} in the $s$-channel is
\begin{equation}
    \label{eq:angularansatz}
    A_{4,s}^{\Delta,J} = \frac{1}{\mathcal{E}} F(S) P_J \bigg( \frac{U-T}{S} \bigg) \,,
\end{equation}
with $P_J(x)$ the $J$-th Legendre polynomial. Again, the intuition comes from the known amplitudes for four massless scalars, where $t$ and $u$ appear in the same way inside the Legendre polynomials. 

\subsection{Differential Equation}
We must now require that the action of the Casimir collapses the exchange to a different, spin-dependent contact interaction. The scattering amplitudes of massless scalars exchanging a massive spin-$J$ particle satisfy
\begin{equation}
    \label{eq:scatteq}
    \Big[ -s+ M^2 \Big] \, \mathcal{A}_{\rm s} = s^J P_J \bigg( \frac{u-t}{s} \bigg) \,.
\end{equation}
For example, one can easily verify that, in the case of a spin-$1$ boson, the mass-dependent part of the propagator vanishes, and the above equation is satisfied.
The natural uplift to a differential equation in the Grassmannian is then:
\begin{equation}
    \label{eq:spineqGrass}
    \Big[ -\mathcal{C}_{12} + (\Delta(3-\Delta) - J(J+1)) \Big] \Bigg[ \frac{1}{\mathcal{E}} F(S) P_J \bigg( \frac{U-T}{S} \bigg) \Bigg] = \frac{\Gamma(J+1)^2}{\mathcal{E}} (2S)^J P_J \bigg( \frac{U-T}{S} \bigg) \,.
\end{equation}
The right-hand side of the above differential equation can be fixed using the known form of $A_{4,s}^{\Delta,J}$ for $\Delta = J+1$, as it does not depend on the mass of the exchanged particle. In particular, we have inferred it from the exchange of photons and gravitons in \cite{Arundine:2026fbr}. For any $J$, one can alternatively identify the source as the contact diagram that precisely recovers the right-hand side of \eqref{eq:scatteq} in the flat-space limit described in~\cite{Arundine:2026fbr}. As a cross-check, we have also confirmed that the correlator it generates in momentum space by encircling the pole in $\tau = 0$ matches the expected result, i.e.~the output of the spin-raising operators in~\cite{Arkani-Hamed:2018kmz, Baumann:2019oyu} on the scalar seed $1/E$. This check was carried out for $J=1$ and $J=2$, thus confirming the intuitive picture. Indeed, the right-hand side of \eqref{eq:firstdiffeq} is also the same for all $\Delta$.

\vskip 4pt
The action of $\mathcal{C}_{12}$ on the ansatz is:
\be
    \mathcal{C}_{12}\Bigg[ \frac{1}{\mathcal{E}} F(S) P_J \bigg( \frac{U-T}{S} \bigg) \Bigg] = \frac{1}{\mathcal{E}} P_J \bigg( \frac{U-T}{S} \bigg) \, \hat{\mathcal{C}}_J \Big[F(S) \Big] \,,
\ee
where we have defined the differential operator
\be
    \hat{\mathcal{C}}_J \Big[F(S) \Big] = \Big[ S^2(2S-1)F''(S) + 2S(3S-1) F'(S) + (2(S+1)-J(J+1)) F(S) \Big] \,.
\ee
We have simplified the result via the recursion relation satisfied by Legendre polynomials
\begin{equation}
    (J+2)P_{J+2}(z) = (2J+3)z P_{J+1}(z) -(J+1)P_J(z) \,.
\end{equation}
Remarkably, the Legendre polynomial factors out of both sides of \eqref{eq:spineqGrass}. It is extraordinary for the Casimir to act this elegantly on the functions we are interested in, given its general form \eqref{eq:terribleCasimir}; it would be interesting to better understand why.

\subsection{Full Solution}
We are left with the following equation to solve:
\begin{equation}
    \label{eq:spinningdiffeq}
    S^2(2S-1) F''(S) + 2S(3S-1) F'(S) + (2(S+1)-\Delta(3-\Delta))F(S) = -\Gamma(J+1)^2 (2S)^J \,.
\end{equation}
Setting $J = 0$ consistently reduces to the scalar case. It is now convenient to define $F(S) = \Gamma(J+1)^2 (2S)^J G(S)$ and solve for $G(S)$. Since only the source depends on $J$, the homogeneous solutions for $F(S)$ are the same as in the scalar case. On the other hand, for generic values of $\Delta$, we have
\begin{equation}
    G(S) = \frac{1}{(\Delta+J-1)(\bar\Delta+J-1)} \, {}_3F_2\Bigg[\begin{array}{c} 1,J+1,J+1\\[ 2pt] \Delta+J, \bar\Delta+J\end{array}\Bigg| \, 2S \Bigg] \,.
\end{equation}
Putting everything together, the correlator of four conformally coupled scalars exchanging a massive spin-$J$ particle is described in the Grassmannian (up to homogeneous pieces) by:
\begin{equation}
    A_{4,s}^{\Delta,J} \supset \frac{\Gamma(J+1)^2}{(\Delta+J-1)(\bar\Delta+J-1)} \frac{1}{\mathcal{E}} \, {}_3F_2\Bigg[\begin{array}{c} 1,J+1,J+1\\[ 2pt] \Delta+J, \bar\Delta+J\end{array}\Bigg| \, 2S \Bigg] (2S)^J P_J \bigg( \frac{U-T}{S} \bigg) \,.
\end{equation}
Note that these functions have a regular $S \to 0$ limit and a logarithmic singularity for $U+T-S = 1-2S =0$, which we must still take care of through the homogeneous solutions. 

\vskip 4pt
The most general solution to \eqref{eq:spineqGrass} takes the form
\begin{align}
\label{eq:scalardospin}
    \boxed{\begin{aligned}
    A_{4,s}^{\Delta,J} & = \frac{\Gamma(J+1)^2}{(\Delta+J-1)(\bar\Delta+J-1)} \frac{1}{\mathcal{E}} \, (2S)^J P_J \bigg( \frac{U-T}{S} \bigg) \Bigg({}_3F_2\Bigg[\begin{array}{c} 1,J+1,J+1\\[2pt] \Delta+J, \bar\Delta+J\end{array}\Bigg| \, 2S \Bigg] \\[4pt]
    & + \frac{c_1(\Delta)}{(2S)^{2-\Delta+J}} \, {}_2F_1\Bigg[\begin{array}{c} \Delta-1, \Delta-1\\[2pt] 2\Delta-2 \end{array}\Bigg| \, 2S \Bigg] + \frac{c_2(\Delta)}{(2S)^{2-\bar\Delta+J}} \, {}_2F_1\Bigg[\begin{array}{c} \bar\Delta-1, \bar\Delta-1\\[2pt] 2\bar\Delta-2 \end{array}\Bigg| \, 2S \Bigg] \Bigg) \,.
    \end{aligned}}
\end{align}
Again, imposing the absence of a branch cut for $S > 1/2$ implies
\be
\begin{aligned}
    \label{eq:c1spinningf}
    c_1(\Delta) & = -\frac{\Gamma(\Delta+J)\Gamma(\bar\Delta+J)\Gamma(\Delta-1)^2}{2 \Gamma(J+1)^2 \Gamma(2\Delta-2)} \Big[ 1 + \tilde f(\Delta) \Big] \,, \\[4pt]
    c_2(\Delta) & = -\frac{\Gamma(\Delta+J)\Gamma(\bar\Delta+J)\Gamma(\bar\Delta-1)^2}{2 \Gamma(J+1)^2 \Gamma(2\bar\Delta-2)} \Big[ 1 - \tilde f(\Delta) \Big] \,.
\end{aligned}
\ee
The AdS$_4$ result and, consequently, the wavefunction coefficient follow from imposing $\tilde f(\Delta) = 1$. In the case of the correlator, we leverage the knowledge of the collapsed limit of the exchange and the result \eqref{eq:scalarftilde}. The combination of these elements suggests the following:
\be
    \tilde f(\Delta)\big|_{J > 0} = \frac{1}{\sin(\pi(\Delta-3/2))} \,,
\ee
i.e.~$A_{4,s}^{\Delta,J}$ is shadow-symmetric, as expected. We will confirm this result in Section~\ref{ssec:spinmomentum}. Again, AdS$_4$ differs via an overall minus sign of $A_{4,s}^{\Delta,J}$. A naive application of the analytic continuation in~\cite{Sleight:2021plv} yields $\tilde f(\Delta) = \csc(\pi(\Delta+J-3/2))$, which is incompatible with the check presented in Section~\ref{ssec:spinmomentum} for odd $J$ and, consequently, with the results of~\cite{Arkani-Hamed:2018kmz, Baumann:2019oyu}. It would be interesting to address this mismatch, as the two prescriptions imply a different collapsed limit.

\subsection{Limiting Cases}
\label{ssec:limitingcases}
In this subsection, we consider the massless limit $\Delta \to J+1$ of the wavefunction coefficient \eqref{eq:scalardospin}, with $J > 0$, together with the partially massless limit $\Delta \to J$ for the case of the graviton. These particles lie neither on the principal ($\Delta = 3/2+i\lambda$) nor on the complementary ($1 < \Delta < 2$) series. We will not discuss the same extrapolation for correlators, where gauge transformations act in a non-trivial way and need to be carefully accounted for~\cite{Ankur:2026ylr}.

\vskip 4pt
For these limiting cases, we have the following simplification
\be
    {}_2F_1\Bigg[\begin{array}{c} \bar\Delta-1, \bar\Delta-1\\[2pt] 2\bar\Delta-2 \end{array}\Bigg| \, 2S \Bigg] \mapsto {}_2F_1\Bigg[\begin{array}{c} -n, -n\\[2pt] -2n \end{array}\Bigg| \, 2S \Bigg] = \mathrm{Poly}_n(2S) \,,
\ee
with $\mathrm{Poly}_n(2S)$ a polynomial of degree $n$. The associated homogeneous solution is therefore a holomorphic function with a $(2S)^{-(\Delta-1+J)}$ pole that does not contribute to the removal of the branch cut for $S > 1/2$.\footnote{This can also be seen from \eqref{eq:appregular}, where the coefficient multiplying $c_2(\Delta)$ vanishes.} As was the case for massless and conformally coupled scalars, then, the choice $\tilde f(\Delta) = 1$ is equivalent to demanding a softer behavior in the $S \to 0$ limit. The same reasoning applies to partially massless particles of any depth.

\vskip 4pt
For $J > 0$, the $\Delta \to J+1$ limit of \eqref{eq:scalardospin} can be extracted with ease:
\be
    A_{4,s}^{J+1,J} = -\frac{\Gamma(J)^2}{2 \mathcal{E} S} \, (2S)^J P_J \bigg( \frac{U-T}{S} \bigg) = \frac{\Gamma(J)^2}{\mathcal{E}^J} \, \tilde S^{J-1} P_J \bigg( \frac{\tilde T- \tilde U}{\tilde S} \bigg) \,,
\ee
which matches the results of \cite{Arundine:2026fbr}, up to a factor arising from the normalization of the right-hand side of \eqref{eq:spineqGrass}. For $J = 0$, the divergent normalization hides the more complicated structure of the $\Delta = 2$ scalar exchange \eqref{eq:scalarD2}, which must be extracted via a more careful procedure.

\vskip 4pt
To describe the exchange of a partially massless graviton, we set $\Delta = J = 2$, thus obtaining
\be
    A_{4,s}^{\rm PM} = -\frac{2(1+S)}{\mathcal{E}} \,P_2\bigg( \frac{U-T}{S} \bigg) \,,
\ee
which is proportional to the graviton answer, times an extra factor of $1+1/S$. It is straightforward to check that summing the residues at $\tau = 0$ and $\tau = \bar \tau_s$, as shown in Figure~\ref{fig:contour} (there is no branch cut in this case), yields the known momentum-space answer obtained from spin-raising the exchange of a scalar with $\Delta = 2$~\cite{Arkani-Hamed:2018kmz}. An interesting feature is that the pole in $S = 0$ is \textit{not simple}! This observation shows that higher-order poles are allowed in the Grassmannian. It would be interesting to better understand the implications of this fact.

\vskip 4pt
We could have solved \eqref{eq:spinningdiffeq} with $\Delta = J+1$ directly. The action of $[-\mathcal{C}_{12} + (\Delta(3-\Delta)-J(J+1))]$ on $A_{4,s}^{J+1,J}$ amounts to a simple multiplication by $-2J^2S$. This feature of the operator and its action on $1/\mathcal{E}$ are what most closely tie the Grassmannian to ordinary flat momentum space. This connection deserves a much more detailed study.

\section{Conversion to Momentum Space}
\label{sec:conversion}
The above results can be converted to momentum space via the integral \eqref{eq:simpleint}, performed along the contour in Figure~\ref{fig:contour}. We omit the momentum-conserving delta function throughout this section. All our solutions generically feature a branch cut for $S < 0$, which, in the $\tau$ plane, corresponds to a branch cut for $0 < \tau < \bar\tau_s$ and one for $\tau > \tau_s$, regulated via $\tilde S \mapsto \tilde S+i\epsilon$ for each Mandelstam. Below, we discuss a few non-trivial examples of the integral \eqref{eq:simpleint}.

\subsection{Conformally Coupled and Massless Scalars}
\label{ssec:scalarcheck}
As a first example, we study the Grassmannian formulae \eqref{eq:scalarD2} and \eqref{eq:scalarD3} for the exchange of conformally coupled and massless scalars, respectively. For both of these cases, when performing the integral \eqref{eq:simpleint} along Figure~\ref{fig:contour}, it is straightforward to check that the semi-arcs close to $\tau = 0$ and $\tau = \bar\tau_s$ are negligible, so that we must simply evaluate
\be
    \psi_{4,s}^{\Delta} = -\int_0^{\bar \tau_s} \frac{\dif \tau}{2 \pi i} \, \mathrm{Disc}_\tau [A_{4,s}^{\Delta}(c_{ij}(\tau))] \,.
\ee
Using the following discontinuities:
\be
\begin{aligned}
    \mathrm{Disc}_{0<\tau<\bar\tau_s}[\mathrm{Li}_2(1-2S(\tau))] & = -2\pi i \log(1-2S(\tau)) \,, \\[4pt]
    \mathrm{Disc}_{0<\tau<\bar\tau_s}[\log 2S(\tau)] & = 2\pi i \,,
\end{aligned}
\ee
we obtain for conformally coupled scalars
\be
    \psi_{4,s}^{\Delta=2} = -\frac{1}{E} \int_0^{\bar\tau_s} \dif\tau \: \frac{\log(1+2|S(\tau)|)}{2\tau |S(\tau)|} \,,
\ee
which matches Equation~(4.10) in~\cite{Goodhew:2021oqg} after integration.

\vskip 4pt
In the case of massless scalars, we find
\be
\psi_{4,s}^{\Delta=3} = -\frac{1}{E} \int_0^{\bar\tau_s} \dif\tau \: \bigg[ \frac{\log(1+2|S(\tau)|)}{2\tau |S(\tau)|} \bigg( 1 + \frac{1}{|S(\tau)|} \bigg)  - \frac{1}{\tau |S(\tau)|} \bigg] \,,
\ee
which matches the action of Equation~(4.58) in~\cite{Arkani-Hamed:2018kmz} on the $\Delta = 2$ wavefunction coefficient after integration.

\vskip 4pt
Both results were checked by numerically comparing the above integrals to the known formulae. Remarkably, we have found the correct results for \textit{wavefunction coefficients} by demanding a soft $S \to 0$ limit in the Grassmannian.

\subsection{Generic Massive Scalars}
As a second example, we consider the Grassmannian formula \eqref{eq:arbitraryscalar} for the exchange \textit{correlator} of a generic massive scalar and show that it matches the known momentum-space answer for the given choice of coefficients $c^{(\rm dS)}_{1,2}(\Delta)$. We first focus on the leading non-analytic term of the exchange in the \textit{collapsed limit} $k_s \to 0$, which is given in the case of $\Delta = 3/2+i \lambda$ by \cite{Arkani-Hamed:2015bza, Arkani-Hamed:2018kmz}
\beq
    \label{eq:collapse}
    \lim_{u,v \to 0} \hat F(u,v) = \Big( \frac{uv}{4} \Big)^{1/2+ i\lambda} \frac{\Gamma \big( \frac{1}{2} + i\lambda \big)^2 \Gamma( -i \lambda)^2}{2\pi} (1 + i \sinh (\pi \lambda)) + \mathrm{c.c.} \,,
\eeq
where the following quantities have been defined:
\beq
    \label{eq:uvdef}
    \hat F \equiv k_s \psi_4, \quad u \equiv \frac{k_s}{k_1 + k_2}, \quad v \equiv \frac{k_s}{k_3 + k_4} \,,
\eeq
with $\psi_4$ the in-in correlator. Just like in the previous subsection, we must compute
\beq
    \label{eq:arbitraryschematic}
    \psi_{4,s}^{\Delta} = -\int_0^{\bar \tau_s} \frac{\dif \tau}{2 \pi i} \, \mathrm{Disc}_\tau [A_{4,s}^{\Delta}(c_{ij}(\tau))] + (\mathrm{semi\text{-}arcs}) \,,
\eeq
with $A_{4,s}^\Delta$ determined in \eqref{eq:arbitraryscalar}. The inhomogeneous solution does not contribute to the non-analytic term, so we can neglect it. Intuitively, we are isolating the particle production contributions, which are completely captured by the homogeneous solutions. This limit is therefore the ideal kinematic setup to confirm the validity of $c_{1,2}(\Delta)$. In Appendix \ref{app:badcomp}, we compute the leading non-analyticity of the above integral, and show that we correctly recover \eqref{eq:collapse}. We also determine the following closed form for the full result, which we explicitly match to the power series representation in \cite{Arkani-Hamed:2018kmz}:
\be
    \label{eq:fullscalarmom}
    \hat F(u,v) = -\frac{1}{2}\frac{u v}{u+v} \bigg[ \frac{\Gamma(\Delta-1)^2}{\Gamma(2\Delta-2)} \bigg( 1 + \frac{1}{\sin(\pi(\Delta-3/2))} \bigg) I_\Delta(u,v) + (\Delta \leftrightarrow\bar\Delta) \bigg] \,,
\ee
where $I_\Delta(u,v)$ is defined in \eqref{eq:uglyKdFdef} for $\Delta = 3/2+i\lambda$. This formula is valid for all $u$ and $v$, is manifestly symmetric in the two variables, and admits an analytic continuation to complex kinematics.

\subsection{Generic Spinning Particles}
\label{ssec:spinmomentum}
As a final example, we study the Grassmannian formula \eqref{eq:scalardospin} for the exchange \textit{correlator} of a generic spinning particle and illustrate how we have checked that the known momentum-space answer~\cite{Arkani-Hamed:2018kmz, Baumann:2019oyu} is correctly recovered.

\vskip 4pt
Unlike the exchange of a scalar, the semi-arc contribution near $\tau = \bar\tau_s$ cannot be neglected, and indeed the first term in \eqref{eq:arbitraryschematic} would be divergent on its own near $\tau = \bar\tau_s$. We therefore adopt a different strategy. We encircle the branch cut connecting $0$ and $\bar\tau_s$ by deforming our integration contour to be the circle $|\tau| = (\bar\tau_s + \tau_s)/2$, so that we have to evaluate
\be
    \psi_{4,s}^{\Delta,J} = \int_{-\pi}^\pi \frac{\dif \theta}{2\pi} \, \frac{\bar\tau_s + \tau_s}{2} \, e^{i \theta} A_{4,s}^{\Delta,J} \Big(c_{ij} \Big( \frac{\bar\tau_s + \tau_s}{2} \, e^{i \theta} \Big) \Big) \,.
\ee
On this contour, the $_3F_2$ term gives a finite contribution and must not be discarded. This integral can be computed numerically for different values of spacelike boundary momenta $k_i^\mu$, and then compared with the value of the known formulae for the same kinematics. We have confirmed that the two outputs match for multiple choices of random kinematics for $J=1$ and $J = 2$, thus further establishing the validity of our chosen coefficients $c^{(\rm dS)}_{1,2}(\Delta)$.

\subsection{Full Solution from Homogeneous Seeds}
\begin{figure}[t!]
	\centering
	\includegraphics[scale=1.19]{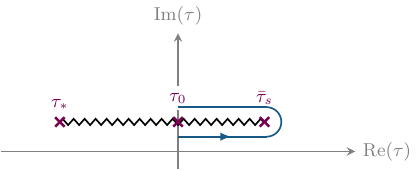} 
	\caption{Effective integration contour for \eqref{eq:opencontour}, when interpreted in terms of the homogeneous solutions only. Their analytic structure differs from that of the full $A_{4,s}$, so the only way to mimic the full result is to have a contour with two endpoints. We have omitted the branch cut below the real line.}
	\label{fig:contour2}
\end{figure}

Throughout the paper, we have treated $A_4$ as a single object with a given analytic structure. However, it is also worthwhile to consider the integral of \eqref{eq:arbitraryscalar} (and similarly for spinning exchanges) over the $\tau$ plane by separating the contributions from the homogeneous and inhomogeneous terms.
\begin{itemize}
    \item The homogeneous terms have a branch cut for both $S(\tau) > 1/2$ and $S(\tau) < 0$. After assigning an $i \epsilon$ prescription to the Mandelstams, these functions have a branch cut for $\tau_* < \tau < \bar \tau_s$ above the real line, where $\tau_*$ is the negative solution of $S(\tau_*) = 1/2$ closest to the origin.
    \item The inhomogeneous term has a branch cut for $S(\tau) > 1/2$, which translates to a branch cut for $\tau_* < \tau< 0$ above the real line.
\end{itemize}
The general contour of integration in Figure~\ref{fig:contour} can be deformed to compute the discontinuity of each term along the respective branch cut above the real line, up to semi-arc contributions around the branch points (which are vanishing for internal scalars).
In particular, we have
\be
\begin{aligned}
    \psi_{4, \rm h} & \propto -\int_{\tau_*}^{\bar \tau_s} \frac{\dif \tau}{2 \pi i} \, \frac{c_1}{\tau E} \, \mathrm{Disc} \Bigg[(2S(\tau))^{\Delta-2}{}_2F_1\Bigg[\begin{array}{c} \Delta-1,\Delta-1\\[2pt] 2\Delta-2\end{array}\Bigg| \, 2S(\tau) \Bigg] \Bigg] + \cdots \,, \\[4pt]
    \psi_{4,\rm inh} & \propto -\int_{\tau_*}^0 \frac{\dif \tau}{2 \pi i} \, \frac{1}{\tau E} \, \mathrm{Disc}_\tau \Bigg[{}_3F_2\Bigg[\begin{array}{c} 1,1,1\\[2pt] \Delta, \bar\Delta\end{array}\Bigg| \, 2S(\tau) \Bigg] \Bigg] \,,
\end{aligned}
\ee
where the ellipses denote the other homogeneous solution, obtained from interchanging $c_1 \leftrightarrow c_2$ and $\Delta \leftrightarrow 3-\Delta$. Since we have tuned the free coefficients to exactly cancel the discontinuity associated with $S > 1/2$, their sum recovers
\be
    \label{eq:opencontour}
    \psi_{4, \rm tot} \propto -\int_{0}^{\bar \tau_s} \frac{\dif \tau}{2 \pi i} \, \frac{c_1}{\tau E} \, \mathrm{Disc} \Bigg[(2S(\tau))^{\Delta-2}{}_2F_1\Bigg[\begin{array}{c} \Delta-1,\Delta-1\\[2pt] 2\Delta-2\end{array}\Bigg| \, 2S(\tau) \Bigg] \Bigg] + \cdots \,.
\ee
In this range, the discontinuity only affects the power law, so that the inhomogeneous solution is irrelevant. However, this is a representation of the \textit{full} solution, which can equivalently be interpreted as a \textit{homogeneous} solution of the Casimir equation, integrated along the open contour in the $\tau$ plane shown in Figure~\ref{fig:contour2}. In practice, the net effect of the inhomogeneous term, in the conversion to momentum space, is simply to reduce the length of the branch cut, therefore modifying the effective integration contour.

\vskip 4pt
We expect the following mechanism to explain how the homogeneous solution to the Casimir equation can generate the full solution in momentum space. Given the action of the Casimir on $A_4(C)$, it can be made to act on $\Lambda$ (appearing as $\delta(C \cdot \Lambda)$ in \eqref{eq:masterint}) by integrating by parts. These steps generate boundary terms if the integral is performed over a region of the orthogonal Grassmannian with boundaries, so a contact interaction emerges as the mismatch. Schematically, we get
\be
    \int \dif C \: \delta(C \cdot \Lambda) \, \mathcal{C}_{12}[A_4(C)] = \Box_{12}[\psi_4(\Lambda)] + (\textrm{boundary term}) \,.
\ee
This implies that, for a function $A_4(C)$ that is annihilated by the operator $[\mathcal{C}_{12} - \Delta(3-\Delta)]$
\be
    \Big[ -\Box_{12} + \Delta(3-\Delta) \Big] \psi_4(\Lambda) = (\textrm{boundary term}) = \frac{1}{E} \,.
\ee
Note that the boundary of the region of integration is located at $\tau = 0$, i.e.~the locus in the Grassmannian where $\mathcal{E}$ vanishes and where contact interactions are entirely localized.

\section{Conclusions}
\label{sec:conclusions}
The cosmological Grassmannian provides a kinematic arena in which de Sitter isometries are manifest and many properties of cosmological correlators resemble those of scattering amplitudes. In this paper, we studied the case of generic massive exchange, where the correlator in momentum space is a rather complicated function of momentum ratios. We managed to find a much simpler function in the Grassmannian, involving hypergeometric (for the massive exchange piece) and Legendre functions (encoding spin information). The result showcases the power of finding a convenient set of kinematic variables for cosmology. Other interesting examples were discussed in~\cite{De:2026shn, Huang:2026tsh, Bala:2026hdm, Bala:2026bdx}. Promising directions for future work include:
\begin{itemize}
    \item It would be interesting to understand how weight-shifting operators act in the Grassmannian, and use them to establish the connection between exchanges of particles with integer-spaced scaling dimensions. 
    \item We have solved the case of four external particles and a single exchange. In principle, the Grassmannian can be used to represent correlators of any number of external particles. In those cases, where several massive exchanges take place, multiple Casimir equations might be necessary to find the solution. Formulating the problem in this space could yield interesting results and potentially make contact with the rules of the \textit{kinematic flow} \cite{Arkani-Hamed:2023kig, Baumann:2024mvm, Baumann:2025qjx, Baumann:2026atn}.
    \item Finally, it would be intriguing to determine expressions for loop diagrams in the Grassmannian, as they will likely be much simpler than their counterparts in momentum space. 
\end{itemize}
With the cleaner description of the simplest tree-level exchange in the Grassmannian at hand, more complicated dynamical questions in cosmology are now within reach. As we better understand the cosmological Grassmannian, the bootstrap rules needed to answer these questions should become clearer. This opens a promising direction for the study of cosmological correlators.

\paragraph{Acknowledgements}
We thank Daniel Baumann, Harry Goodhew, Mang Hei Gordon Lee, Andrzej Pokraka, Facundo Rost, Kamran Salehi Vaziri and Tom Westerdijk for insightful discussions. 

\vskip 4pt
The research of MA is funded by the European Union (ERC,  \raisebox{-2pt}{\includegraphics[height=0.9\baselineskip]{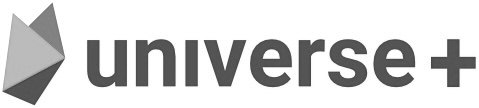}}, 101118787). GLP is supported by the ERC (NOTIMEFORCOSMO, 101126304), by Scuola Normale, and by INFN (IS GSS-Pi). GLP is also supported by the Italian Ministry of Universities and Research (MUR) under contract 20223ANFHR (PRIN2022).

\newpage
\appendix

\section{Determination of the Free Coefficients}
\label{app:disc}
In this appendix, we prove that the absence of a branch cut for $S > 1/2$ in \eqref{eq:scalardospin} implies the result \eqref{eq:c1spinningf}. After setting up the general scenario, we will show the explicit computations for $J = 0$, and argue how they generalize to higher spin.

\subsection{Branch Cut Removal}
We recall the discontinuity of the hypergeometric ${}_2F_1$
\begin{align}
    \label{eq:app2F1disc}
    \mathrm{Disc}_{z>1}\Bigg[{}_2F_1\Bigg[\begin{array}{c}a,b\\[2pt] c \end{array}\Bigg| \, z \Bigg] \Bigg] & \equiv {}_2F_1\Bigg[\begin{array}{c}a,b\\[2pt] c \end{array}\Bigg| \, z+i\epsilon \Bigg] - {}_2F_1\Bigg[\begin{array}{c}a,b\\[2pt] c \end{array}\Bigg| \, z-i\epsilon \Bigg] \nonumber\\[4pt]
    & = \frac{2\pi i \,\Gamma(c) (z-1)^{c-a-b}}{\Gamma(a)\Gamma(b)\Gamma(c-a-b+1)} \, {}_2F_1\Bigg[\begin{array}{c}c-a,c-b\\[2pt] c-a-b+1 \end{array}\Bigg| \, 1-z \Bigg] \,.
\end{align}
Given the integral representation of the hypergeometric $_{3}F_2$
\be
    {}_3F_2\Bigg[\begin{array}{c}a_1, a_2, a_3 \\[2pt] b_1, b_2 \end{array}\Bigg| \, z \Bigg] = \frac{\Gamma(b_1)}{\Gamma(a_1)\Gamma(b_1-a_1)} \int_0^1 \dif t \: t^{a_1-1} (1-t)^{b_1-a_1-1} \, {}_2F_1\Bigg[\begin{array}{c}a_2, a_3\\[2pt] b_2 \end{array}\Bigg| \, t z \Bigg] \,,
\ee
we can take its discontinuity for the special choice of parameters appearing in \eqref{eq:scalardospin}
\be
    \mathrm{Disc}_{z>1}\Bigg[ {}_3F_2\Bigg[\begin{array}{c} 1, \hs J+1, \hs J+1 \\[2pt] \Delta+J, \hs 3-\Delta+J \end{array}\Bigg| \, z \Bigg] \Bigg] = (\Delta+J-1)\int_{1/z}^1 \dif t \: (1-t)^{\Delta+J-2} \, \mathcal{F}(\Delta,J,tz) \,,
\ee
where we have defined
\be
    \mathcal{F}(\Delta,J,tz) \equiv \frac{2\pi i \, \Gamma(3-\Delta+J)(tz-1)^{1-\Delta-J}}{\Gamma(J+1)^2 \Gamma(2-\Delta-J)} \, {}_2F_1\Bigg[\begin{array}{c} 2-\Delta, 2-\Delta \\[2pt] 2-\Delta-J \end{array}\Bigg| \, 1-tz \Bigg]\,.
\ee
The hypergeometric function appearing above takes the simple form
\be
    {}_2F_1\Bigg[\begin{array}{c} 2-\Delta, 2-\Delta \\[2pt] 2-\Delta-J \end{array}\Bigg| \, 1-tz \Bigg] = (t z)^{\Delta-2-J} Q_J(1-t z) \,,
\ee
with $Q_J$ a polynomial of degree $J$, which allows us to express the result of the integral in terms of a sum of ${}_2F_1$ functions. In the special case of $J=0$, we have $Q_0 = 1$ and
\be
    \label{eq:app3F2disc}
    \mathrm{Disc}_{z>1}\Bigg[ {}_3F_2\Bigg[\begin{array}{c} 1, 1, 1 \\[2pt] \Delta, 3-\Delta \end{array}\Bigg| \, z \Bigg] \Bigg] = \frac{2\pi^2 i (\Delta-1)(\Delta-2)}{\sin(\pi\Delta)} \frac{1}{z} \, {}_2F_1\Bigg[\begin{array}{c} \Delta-1, \hs 2-\Delta \\[2pt] 1 \end{array}\Bigg| \, 1-\frac{1}{z} \Bigg] \,.
\ee
Substituting the discontinuities \eqref{eq:app2F1disc} and \eqref{eq:app3F2disc} in \eqref{eq:firstsol} yields
\begin{align}
    \label{eq:scalardisc}
    \frac{(\Delta-1)(2-\Delta)}{2\pi i} \hs \mathrm{Disc}_{S>1/2}[F(S)] & = \frac{\pi(\Delta-1)(\Delta-2)}{\sin(\pi\Delta)} \frac{1}{2S} \,{}_2F_1\Bigg[\begin{array}{c} \Delta-1, 2-\Delta \\[2pt] 1 \end{array}\Bigg| \, 1-\frac{1}{2S} \Bigg] \nonumber\\[4pt]
    & + c_1(\Delta) \frac{\Gamma(2\Delta-2)}{\Gamma(\Delta-1)^2} (2S)^{\Delta-2} \, {}_2F_1\Bigg[\begin{array}{c} \Delta-1, \Delta-1 \\[2pt] 1 \end{array}\Bigg| \, 1-2S \Bigg] \nonumber\\[4pt]
    & + c_2(\Delta) \frac{\Gamma(4-2\Delta)}{\Gamma(2-\Delta)^2} (2S)^{1-\Delta} \, {}_2F_1\Bigg[\begin{array}{c} 2-\Delta, 2-\Delta \\[2pt] 1 \end{array}\Bigg| \, 1-2S \Bigg] \,.
\end{align}
Substituting the following identities into the expression above
\be
\begin{aligned}
    {}_2F_1\Bigg[\begin{array}{c} \Delta-1, \Delta-1 \\[2pt] 1 \end{array}\Bigg| \, 1-2S \Bigg] & = (2S)^{1-\Delta} \, {}_2F_1\Bigg[\begin{array}{c} \Delta-1, 2-\Delta \\[2pt] 1 \end{array}\Bigg| \, 1- \frac{1}{2S} \Bigg] \,,\\
    {}_2F_1\Bigg[\begin{array}{c} 2-\Delta, 2-\Delta \\[2pt] 1 \end{array}\Bigg| \, 1-2S \Bigg] & = (2S)^{\Delta-2} \, {}_2F_1\Bigg[\begin{array}{c} \Delta-1, 2-\Delta \\[2pt] 1 \end{array}\Bigg| \, 1- \frac{1}{2S} \Bigg] \,,
\end{aligned}
\ee
implies that $\mathrm{Disc}_{S>1/2}[F(S)]$ is set to $0$ by demanding
\be
    \label{eq:appregular}
    \Gamma(\Delta) \Gamma(3-\Delta) + c_1(\Delta) \frac{\Gamma(2\Delta-2)}{\Gamma(\Delta-1)^2} + c_2(\Delta) \frac{\Gamma(4-2\Delta)}{\Gamma(2-\Delta)^2} = 0 \,.
\ee
It is then straightforward to check that \eqref{eq:generalc1Del} is a parametrization of the solutions to the above condition. The AdS$_4$ solution then easily follows from further imposing $c_2(\Delta) = 0$ in \eqref{eq:appregular}.

\vskip 4pt
The $J>0$ case easily follows from the identity
\be
    {}_3F_2\Bigg[\begin{array}{c} 1,J+1,J+1\\[2pt] \Delta+J, \bar\Delta+J\end{array}\Bigg| \, 2S \Bigg] = \frac{\Gamma(\Delta+J)\Gamma(\bar\Delta+J)}{\Gamma(\Delta)\Gamma(\bar\Delta)\Gamma(J+1)^2} \, \frac{1}{(2S)^J} \Bigg( {}_3F_2\Bigg[\begin{array}{c} 1,1,1\\[2pt] \Delta, \bar\Delta\end{array}\Bigg| \, 2S \Bigg] + \cdots \Bigg) \,,
\ee
where the neglected term is a polynomial of degree $J-1$ that does not contribute to the discontinuity for $S > 1/2$. This relation, together with \eqref{eq:generalc1Del}, automatically returns \eqref{eq:c1spinningf}.  

\subsection{Branch Point Regularization}
A simpler way to obtain \eqref{eq:c1spinningf} involves studying the local behavior of $F(S)$ close to $S = 1/2$. Although this does not imply, in principle, the absence of a discontinuity for all $S > 1/2$, it is a necessary condition that yields the same result. For $J = 0$, we find
\begin{equation}
    \begin{aligned}
        F(S \to 1/2) & \simeq \mathcal{P}(\Delta,c_1,c_2) \log(1-2S) + \mathcal{O}(1) \,, \\
        (\Delta-1)(\Delta-2) \,\mathcal{P}(\Delta,c_1,c_2) & = \Gamma(\Delta) \Gamma(3-\Delta) + c_1(\Delta) \frac{\Gamma(2\Delta-2)}{\Gamma(\Delta-1)^2} + c_2(\Delta) \frac{\Gamma(4-2\Delta)}{\Gamma(2-\Delta)^2} \,.
    \end{aligned}
\end{equation}
For arbitrary choices of the free coefficients, the function has a local logarithmic behavior that is responsible for the branch cut for $S > 1/2$. We can remove it by imposing $\mathcal{P} = 0$, which is equivalent to \eqref{eq:appregular}, and therefore yields the same result. Performing the same study for $J > 0$ then easily returns \eqref{eq:c1spinningf}. 

\vskip 4pt
The reason why simply regulating $S = 1/2$ guarantees the entire removal of the branch cut is the following. The discontinuity function---or, more specifically, its analytic continuation away from the $S > 1/2$ half-line---is always a homogeneous solution of the same Casimir equation as the exchange diagram \eqref{eq:spinningdiffeq}. By imposing that the discontinuity vanishes at $S = 1/2$, we simultaneously enforce the solution that does not have a local logarithmic behavior, and impose its overall coefficient to vanish, therefore removing the branch cut entirely. For example, \eqref{eq:scalardisc} is indeed the homogeneous solution of the Casimir equation without a logarithmic behavior at $S = 1/2$, which is itself non-vanishing at that point.

\newpage
\section{Computation of the Scalar Exchange}
\label{app:badcomp}
In this appendix, we study the integral \eqref{eq:arbitraryschematic}, first in the $k_s \to 0$ limit, then in full generality. 

\subsection{Collapsed Limit}
Before we begin, we recall the following discontinuity:
\be
    \label{eq:xalphadisc}
    \mathrm{Disc}_{x<0} [x^\alpha] \equiv (-|x| + i \epsilon)^\alpha - (-|x| - i \epsilon)^\alpha = 2 i \sin(\pi \alpha) |x|^\alpha \,.
\ee
By defining $x \equiv \langle \bar 1 \bar 2 \rangle \langle \bar 3 \bar 4 \rangle \, \tau/E$, equation \eqref{eq:arbitraryschematic} reduces (for $\Delta = 3/2 + i \lambda$) to
\be
    \psi_{4,s}^\Delta = -\frac{1}{2\pi i} \int_0^{\bar x_s} \frac{\dif x}{x E} \, \frac{c_1}{\lambda^2+1/4} {}_2F_1\Bigg[\begin{array}{c} \frac{1}{2}+i\lambda, \frac{1}{2}+i\lambda \\[2pt] 1+2i\lambda \end{array}\Bigg| \, 2S(x) \Bigg] \, \mathrm{Disc}_x[(2S(x))^{-1/2+i \lambda}] + \cdots \,,
\ee
where we do not report the analogous $c_2$ term and the contribution from the semi-arcs. In terms of $u,v$ defined in \eqref{eq:uvdef}, we have
\be
    \label{eq:xsxbsdef}
    2S(x) = - \frac{(x-\bar x_s)(x-x_s)}{x}, \quad \bar x_s = \frac{uv(1-u)(1-v)}{(u+v)^2}, \quad x_s = \frac{uv(1+u)(1+v)}{(u+v)^2} \,.
\ee
We recall the surprising feature that, for this range of $\tau$, the ${}_3F_2$ has no discontinuity and does not contribute to the integral, although we are describing the \textit{full} solution in momentum space. We observe that the discontinuity does not affect the hypergeometric functions, since $2S(x)$ ranges from $-\infty$ to $0$. In particular, using \eqref{eq:xalphadisc} with respect to the variable $S$ yields
\be
    \mathrm{Disc}_x[(2S(x))^{-1/2+ i \lambda}] = -2i \cosh (\pi \lambda) (2|S(x)|)^{-1/2 + i \lambda} \,,
\ee
where the sign is determined by $S'(x) > 0$ in the relevant region of integration.

\vskip 4pt
We first argue that the contribution from the semi-arcs vanishes. Near $x = \bar x_s$, the integrand behaves as $(x-\bar x_s)^{-1/2 \pm i \lambda}$, i.e.~its integral along the semi-arc tends to $0$. Similarly, near $x = 0$, one can use hypergeometric identities to show that the integrand behaves logarithmically. From now on, we will focus on the $c_1$ term only for clarity:
\be
    \psi_{4,s}^\Delta \supset \frac{c_1 \cosh(\pi \lambda)}{\pi E (\lambda^2 + 1/4)} \int_0^{\bar x_s} \frac{\dif x}{x} \, (2|S(x)|)^{-1/2 + i \lambda} \, {}_2F_1\Bigg[\begin{array}{c} \frac{1}{2}+i\lambda, \frac{1}{2}+i\lambda \\[2pt] 1+2i\lambda \end{array}\Bigg| \, -2|S(x)| \Bigg] \,.
\ee
Owing to the monotonicity of $S(x)$ in $x \in (0, \bar x_s)$, it is convenient to integrate with respect to $2|S(x)| \equiv p(x)$ instead, which returns
\be
    \label{eq:intrewrite}
    \frac{c_1 \cosh(\pi \lambda)}{\pi E (\lambda^2 + 1/4)} \int_0^{+\infty} \frac{\dif p}{\sqrt{(p + x_s + \bar x_s)^2 - 4 x_s \bar x_s}} \, p^{-1/2 + i \lambda} \, {}_2F_1\Bigg[\begin{array}{c} \frac{1}{2}+i\lambda, \frac{1}{2}+i\lambda \\[2pt] 1+2i\lambda \end{array}\Bigg| \, -p \Bigg] \,.
\ee
We now want to study the $k_s \to 0$ limit of the above integral. To this end, we first observe that
\be
    x_s - \bar x_s = \frac{2 u v}{u+v} \equiv \epsilon \ll 1 \,,
\ee
which allows us to rewrite the square root in \eqref{eq:intrewrite} as
\be
    \sqrt{(p + x_s + \bar x_s)^2 - 4 x_s \bar x_s} = \sqrt{(p + \epsilon)^2 + 4 \bar x_s p} \,.
\ee
The \textit{dominant non-analytic term} in $\epsilon$ can be easily extracted via the \textit{method of regions}.

\vskip 6pt
\begin{eBox3}
{\bf Method of Regions}: \ 
We briefly outline the ``method of regions'' for evaluating small parameter expansions of integrals. Consider the following function of $\epsilon$:
\be
    I(\epsilon) = \int_0^\infty \dif x \, F(\epsilon;x) \,.
\ee
We want to study the $\epsilon \to 0$ expansion of $I(\epsilon)$. First, there exists a characteristic scale $\bar x(\epsilon)$ that naturally splits the domain of integration into a ``hard'' region $x \gg \bar x(\epsilon)$ and a ``soft'' region $x \ll \bar x(\epsilon)$. We then split the integral accordingly:
\be
    \int_0^{+\infty} \dif x \, F(\epsilon;x) = \int_0^{\bar x(\epsilon)} \dif x \, F(\epsilon;x) + \int_{\bar x(\epsilon)}^{+\infty} \dif x \, F(\epsilon;x) \,.
\ee
In the hard region, one performs a series expansion of the integrand $F(\epsilon;x)$ and extends the domain of integration back to $(0,+\infty)$. Similarly, in the soft region, one first changes the variable of integration to $x = \bar x(\epsilon) \, y$, and then performs the same operations:
\be
\begin{aligned}
    \int_{\bar x(\epsilon)}^{+\infty} \dif x \, F(\epsilon;x) & \mapsto \sum_{n=0}^{+\infty} \frac{1}{n!} \int_0^{+\infty} \dif x \, \partial_\epsilon^n \big[F(\epsilon;x) \big] \big|_{\epsilon = 0} \, \epsilon^n \equiv I_{\rm hard}(\epsilon) \,, \\[4pt]
   \int_{0}^{\bar x(\epsilon)} \dif x \, F(\epsilon;x) & \mapsto \sum_{n=0}^{+\infty} \frac{1}{n!} \int_0^{+\infty} \dif y \: \bar x(\epsilon) \, \partial_\epsilon^n \big[ F(\epsilon; \bar x(\epsilon) \, y) \big] \big|_{\epsilon = 0} \, \epsilon^n \equiv I_{\rm soft}(\epsilon) \,. 
\end{aligned}
\ee
Unsurprisingly, the hard region has IR divergences and the soft region has UV divergences, but the sum of the two terms above is the finite expansion of $I(\epsilon) = I_{\rm soft}(\epsilon) + I_{\rm hard}(\epsilon)$. 

\vskip 4pt
Throughout this example, we have assumed that $I(\epsilon)$ is analytic at $\epsilon = 0$, which is not the case for the computation at hand. However, the strategy we have described is still valid in less regular scenarios.
\end{eBox3}

\vspace{0.05cm}
The natural scale that identifies the two regions is the value $\bar p$ found by studying the saddles of the integrand, i.e.
\be
    \bar p \sim \frac{\epsilon^2}{4 \bar x_s} \,.
\ee
It is straightforward to verify that the hard region only contributes to the analytic part of $\psi_{4,s}^\Delta$. We can study the soft region by setting $p = \bar p \, y$:
\be
    \frac{c_1 \cosh(\pi \lambda)}{\pi E (\lambda^2 + 1/4)} \frac{1}{\sqrt{4 \bar x_s}} \bigg( \frac{\epsilon^2}{4 \bar x_s} \bigg)^{i \lambda} \int_0^{+\infty} \frac{\dif y}{\sqrt{y + \big( 1 + \frac{\epsilon y}{4 \bar x_s}\big)^2}} \, y^{-1/2 + i \lambda} \, {}_2 F_1\bigg[\cdots \bigg| -\frac{\epsilon^2 y}{4 \bar x_s}\bigg] \,.
\ee
We find the leading contribution by performing the integral after setting $\epsilon = 0$:
\be
    \psi_{4,s}^\Delta \supset \frac{c_1 \cosh(\pi \lambda)}{\pi E (\lambda^2 + 1/4)} \frac{\Gamma \big( \frac{1}{2} + i \lambda \big) \Gamma(-i \lambda)}{\sqrt{4 \pi \bar x_s}} \bigg( \frac{\epsilon^2}{4 \bar x_s} \bigg)^{i \lambda} \,.
\ee
Substituting $\bar x_s \approx uv/(u+v)^2$ and the general value of $c_{1,2}(\Delta)$ allowed by the absence of folded singularities \eqref{eq:generalc1Del} yields
\begin{align}
    \hat F(u,v) \equiv k_s \psi_{4,s}^\Delta \supset & \: -\frac{(uv)^{1/2+i\lambda}}{\sqrt{4\pi}} \frac{\cosh(\pi\lambda)}{2\pi} \frac{\Gamma \big( \frac{3}{2}+i\lambda \big) \Gamma \big( \frac{3}{2} - i\lambda \big)}{\lambda^2 + 1/4} \nonumber \\[4pt]
    & \times \frac{\Gamma \big( \frac{1}{2} + i \lambda \big)}{\Gamma(1+2i\lambda) \Gamma(-i\lambda)} \Gamma \Big( \frac{1}{2} + i \lambda \Big)^2 \Gamma(-i\lambda)^2 (1 + \tilde f(3/2+i\lambda)) \,.
\end{align}
Using Gamma function identities simplifies the above result to
\be
    \lim_{u,v\to 0} \hat F(u,v) = -\Big( \frac{uv}{4} \Big)^{1/2+ i\lambda} \frac{\Gamma \big( \frac{1}{2} + i\lambda \big)^2 \Gamma( -i \lambda)^2}{2\pi} (-i \sinh(\pi\lambda))(1 + \tilde f(3/2+i\lambda)) + \mathrm{c.c.} \,,
\ee
which matches the desired result \eqref{eq:collapse}, provided that we set
\be
    \tilde f(\Delta) = \frac{1}{\sin (\pi(\Delta-3/2))} \,,
\ee
as claimed in Section~\ref{ssec:partprod}. In the computations above, we have implicitly assumed a finite $\bar x_s$ as $u,v \to 0$, i.e.~$u/v = (k_3+k_4)/(k_1+k_2) = \mathcal{O}(1)$, consistently with just demanding $k_s \to 0$.

\vskip 4pt
In AdS$_4$, on the other hand, the bulk-to-bulk propagator in momentum space takes the form \cite{Mueck:1998wkz, Liu:1998ty}
\be
    G_F(z,z') = (z z')^{3/2} I_\nu(k_s z_<) K_\nu(k_s z_>) \,,
\ee
with $\nu = \Delta - 3/2$, $z_< = \mathrm{min}(z,z')$ and $z_> = \mathrm{max}(z,z')$. In the collapsed limit, it is easy to show from the Witten diagram that one has
\be
    \lim_{u,v \to 0} \hat F^{(\rm AdS)}(u,v) = \Big( \frac{uv}{4} \Big)^{1/2+\nu} \frac{\Gamma \big( \frac{1}{2} + \nu \big)^2 \Gamma( -\nu)}{\Gamma(1+\nu)} \,,
\ee
which confirms the boundary condition $c_2^{(\rm AdS)}(\Delta) = 0$ and is correctly predicted by the value of $c_1^{(\rm AdS)}(\Delta)$ in \eqref{eq:scalarc1c2dsAds} imposed by regularity in the Grassmannian. Performing a Wick rotation on the Witten diagram $z \mapsto -i\eta(1-i\epsilon)$ recovers the time integral representation of the wavefunction coefficient, thus also corroborating the validity of $c_{1,2}^{(\rm WF)}(\Delta)$.

\subsection{Full Result}
\begin{figure}[t!]
	\centering
    \includegraphics[scale=0.55]{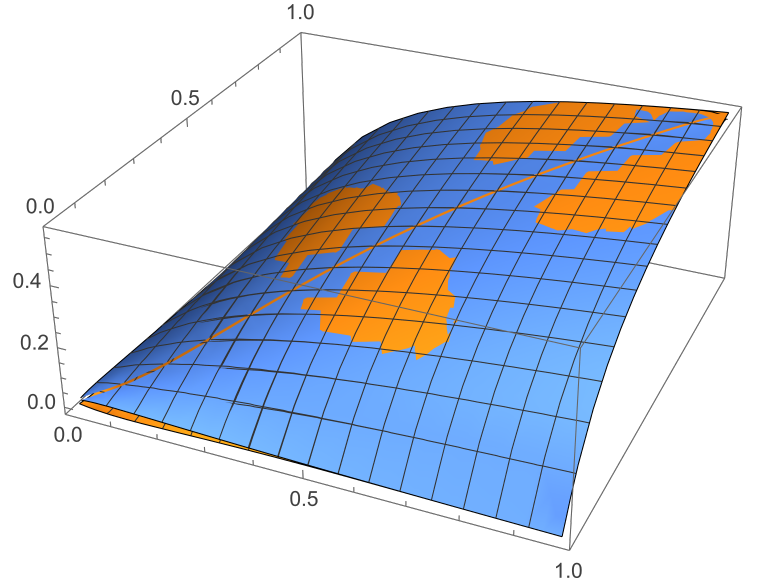} \quad  
	\includegraphics[scale=0.6]{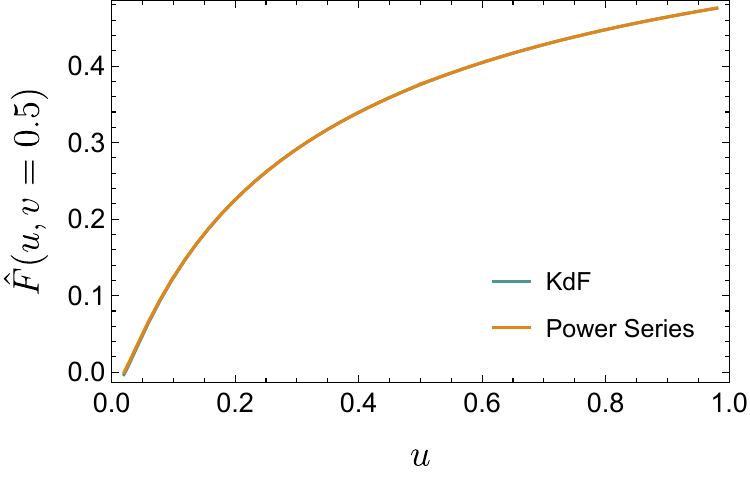} 
	\caption{Numerical comparison between \eqref{eq:fullscalarmom} (in blue) and (3.36) in \cite{Arkani-Hamed:2018kmz} (in orange) for the case of $\Delta = 3/2 + i$. All the sums in the KdF series representation and in (3.36) range from $0$ to $N = 20$. Left: plot in the $(u,v) \in (0,1)^2$ plane. Right: section of the plot for $v = 0.5$.}
	\label{fig:numerimatch}
\end{figure}

We now go back to the study of the integral \eqref{eq:intrewrite}
\be
    I_\Delta(u,v) = \int_0^{+\infty} \frac{\dif p}{\sqrt{(p + x_s + \bar x_s)^2 - 4 x_s \bar x_s}} \, p^{-1/2 + i \lambda} \, {}_2F_1\Bigg[\begin{array}{c} \frac{1}{2}+i\lambda, \frac{1}{2}+i\lambda \\[2pt] 1+2i\lambda \end{array}\Bigg| \, -p \Bigg] \,,
\ee
which we will soon show to be equal to the following combination of \textit{Kampé de Fériet} (KdF) functions:
\begin{align}
    \label{eq:uglyKdFdef}
   I_\Delta(u,v) & = \frac{\pi p_-^{i\lambda}}{p_+^{1/2} \cosh(\pi \lambda)} F^{1|1|1}_{\,0|1|1}\Bigg[\begin{array}{c} \frac{1}{2}+i\lambda\\[2pt]-\end{array}\Bigg| \begin{array}{c} \frac{1}{2} + i\lambda\\[2pt] 1+2i\lambda \, \end{array} \Bigg| \begin{array}{c} \frac{1}{2} \\[2pt] 1 \end{array}\Bigg| \, p_-, 1-\frac{p_-}{p_+}\Bigg] \nonumber \\[4pt]
   & - \frac{1}{\lambda^2+1/4} \frac{\Gamma(1+2i\lambda)}{\Gamma\big( \frac{1}{2} + i \lambda \big)^2} \bigg( \frac{p_-}{p_+} \bigg)^{1/2} F^{1|2|1}_{\,0|2|1}\Bigg[\begin{array}{c} 1 \\[2pt]-\end{array}\Bigg| \begin{array}{c} 1,1\\[2pt] \frac{3}{2}+i\lambda, \frac{3}{2} -i\lambda \, \end{array} \Bigg| \begin{array}{c} \frac{1}{2} \\[2pt] 1 \end{array}\Bigg| \, p_-, 1-\frac{p_-}{p_+}\Bigg] \,,
\end{align}
where $p_\pm \equiv (\sqrt{x_s(u,v)} \pm \sqrt{\bar x_s(u,v)})^2$, and $\bar x_s, x_s$ were defined in \eqref{eq:xsxbsdef}. The KdF functions admit the following power series representations
\be
\begin{aligned}
    \label{eq:KdFseries}
    F^{1|1|1}_{\,0|1|1}\Bigg[\begin{array}{c} a_1\\[2pt] - \end{array}\Bigg| \begin{array}{c} a_2\\[2pt] b_2 \, \end{array} \Bigg| \begin{array}{c} a_3 \\[2pt] b_3 \end{array}\Bigg| \, x, y\Bigg] & \equiv \sum_{m,n=0}^{+\infty} \frac{(a_1)_{m+n} (a_2)_m (a_3)_n}{(b_2)_m (b_3)_n} \frac{x^m}{m!} \frac{y^n}{n!} \,, \\[4pt]
    F^{1|2|1}_{\,0|2|1}\Bigg[\begin{array}{c} a_1\\[2pt] -\end{array}\Bigg| \begin{array}{c} a_2,\tilde a_2\\[2pt] b_2, \tilde b_2 \, \end{array} \Bigg| \begin{array}{c} a_3 \\[2pt] b_3 \end{array}\Bigg| \, x, y\Bigg] & \equiv \sum_{m,n=0}^{+\infty} \frac{(a_1)_{m+n} (a_2)_m (\tilde a_2)_m (a_3)_n}{(b_2)_m (\tilde b_2)_m (b_3)_n} \frac{x^m}{m!} \frac{y^n}{n!} \,.
\end{aligned}
\ee
The first KdF function is, in particular, an Appell $F_2$. The sum of this integral with its shadow \eqref{eq:fullscalarmom} is shown to match the result of \cite{Arkani-Hamed:2018kmz} in Figure~\ref{fig:numerimatch}.

\vskip 4pt
To prove this result, we first substitute the Mellin-Barnes representation of the hypergeometric function
\be
    {}_2F_1\Bigg[\begin{array}{c} \frac{1}{2}+i\lambda, \frac{1}{2}+i\lambda \\[2pt] 1+2i\lambda \end{array}\Bigg| \, -p \Bigg] = \frac{\Gamma(1+2i\lambda)}{\Gamma \big(\frac{1}{2} + i \lambda \big)^2} \int_{-i\infty}^{+i\infty} \frac{\dif s}{2\pi i} \frac{\Gamma \big(\frac{1}{2} + i \lambda +s \big)^2\Gamma(-s)}{\Gamma(1+2i\lambda+s)} \, p^s \,.
\ee
This allows us to easily evaluate the integral with respect to $p$ \cite{Gradshteyn:2007zz}:
\begin{align}
    \frac{p_-^{i\lambda+s}}{p_+^{1/2}} \, \mathcal{I}_\Delta(u,v;s) & \equiv \int_0^{+\infty} \dif p \: \frac{p^{-1/2+i\lambda+s}}{\sqrt{(p+p_-)(p+p_+)}} \nonumber \\[4pt]
    & =\frac{p_-^{i\lambda+s}}{p_+^{1/2}} \Gamma \bigg( \frac{1}{2}+i\lambda+s \bigg) \Gamma \bigg( \frac{1}{2} - i\lambda-s \bigg) {}_2F_1\Bigg[\begin{array}{c} \frac{1}{2}, \frac{1}{2}+i\lambda+s \\[2pt] 1 \end{array}\Bigg| \, 1- \frac{p_-}{p_+} \Bigg] \,.
\end{align}
We must now perform the integral with respect to $s$:
\be
    I_\Delta(u,v) = \frac{p_-^{i\lambda}}{p_+^{1/2}} \frac{\Gamma(1+2i\lambda)}{\Gamma \big(\frac{1}{2} + i \lambda \big)^2} \int_{-i\infty}^{+i\infty} \frac{\dif s}{2\pi i} \frac{\Gamma \big(\frac{1}{2} + i \lambda +s \big)^2 \Gamma(-s)}{\Gamma(1+2i\lambda+s)}  \, (p_-)^s \,\mathcal{I}_\Delta(u,v;s)\,.
\ee
This is achieved by summing the residues of the integrand in $s = m$ and $s = m+1/2-i\lambda$, with $m = 0,1,2,\dots$, arising from the Gamma functions. It is straightforward to show that these two sets of poles, combined with the power series representation of the hypergeometric ${}_2F_1$, reproduce exactly the power series of the two KdF functions with the appropriate prefactors \eqref{eq:uglyKdFdef}. Given this result, the full correlator of the scalar exchange is given by
\begin{align}
    \hat F(u,v) & = \frac{u v}{u+v} \frac{\cos(\pi(\Delta-3/2))}{\pi(\Delta-1)(\bar\Delta-1)} \, [c_1(\Delta) I_\Delta(u,v) + c_2(\Delta) I_{\bar \Delta}(u,v)] \nonumber \\[4pt]
    & = -\frac{1}{2}\frac{u v}{u+v} \bigg[ \frac{\Gamma(\Delta-1)^2}{\Gamma(2\Delta-2)} \bigg( 1 + \frac{1}{\sin(\pi(\Delta-3/2))} \bigg) I_\Delta(u,v) + (\Delta \leftrightarrow\bar\Delta) \bigg] \,.
\end{align}
The correlator in AdS$_4$ is obtained by only keeping the first term in the above formula, with $\sin(\pi(\Delta-3/2)) \mapsto 1$, and flipping the overall sign.

\newpage
\phantomsection
\addcontentsline{toc}{section}{References}
\bibliographystyle{utphys}
{\linespread{1.075}
	\bibliography{ref.bib}
}

\end{document}